\documentclass[12pt,twocolumn]{elsart2}

\usepackage{graphics}
\usepackage{amssymb}

\begin{document}

\begin{frontmatter}

\title{Cosmic rays through the Higgs portal}

\author{Rainer Dick\corauthref{cor1}\thanksref{label1}\thanksref{label2}}
\ead{rainer.dick@usask.ca},
\author{Robert B. Mann\thanksref{label1}\thanksref{label3}}
\ead{rmann@perimeterinstitute.ca},
\author{Kai E. Wunderle\thanksref{label2}}
\ead{kai.wunderle@usask.ca}
\corauth[cor1]{Corresponding author}
\address[label1]{Perimeter Institute for Theoretical Physics,
        31 Caroline Street North, Waterloo,\\ Ontario, Canada N2L 2Y5}
\address[label2]{Department of Physics and Engineering Physics,
        University of Saskatchewan,\\ 
        116 Science Place, Saskatoon, Saskatchewan, Canada S7N 5E2}
\address[label3]{Department of Physics and Astronomy, University of Waterloo,\\
        200 University Avenue West, Waterloo, Ontario, Canada N2L 3G1}

\begin{abstract}

We consider electroweak singlet dark matter with a mass comparable to the
Higgs mass. The singlet is assumed to couple to standard matter through a
perturbative coupling to the Higgs particle. The annihilation of a
singlet in the mass range $m_S\sim m_h$ is dominated by proximity to the 
$W$, $Z$ and Higgs peaks in the annihilation cross section. We find that
the continuous photon spectrum from annihilation of perturbatively 
coupled singlets in the galactic halo can reach a level of several
per mil of the EGRET diffuse $\gamma$ ray flux.
\end{abstract}

\begin{keyword}
Cosmic rays, electroweak singlet, dark matter
\PACS 12.60.Fr \sep 95.30.Cq \sep 95.35.+d \sep 96.50.S- \sep 98.70.Sa
\end{keyword}

\end{frontmatter}

\section{Introduction}\label{sec:intro}

Revealing the distribution and nature of dark matter is
one of the most interesting current challenges of particle
physics and astrophysics. Numerous candidates have been 
proposed and studied, including axions \cite{axion},
neutralinos \cite{neutralino}, Kaluza-Klein photons \cite{rocky1},
Kaluza-Klein or string dilatons \cite{dilaton}, 
and superheavy dark matter either directly from inflationary
expansion \cite{rocky2,KT} or from preheating after inflation \cite{lev}.
Unstable particles like axions and dilatons have to be very light,
in the sub GeV range, to survive long enough to serve as dark matter.
However, neutralinos are usually assumed to have masses beyond 100 GeV,
although lower mass limits strongly depend on supersymmetric models
\cite{dan,dan2,bottino}.

Another very interesting model for dark matter which can have a large 
mass is a stable electroweak singlet $S$ which couples to standard model 
matter exclusively through a coupling to the Higgs boson $H$,
\begin{equation}\label{eq:coupl1}
H_I=\frac{\eta}{2}\int d^3\vec{x} S^2 H^\dagger H.
\end{equation}
We focus in particular on the $Z_2$ symmetric model proposed in 
\cite{zee,mcd,cliff,mura,frank}, which also allows for a bare mass 
term for the singlet, but does not include Higgs-singlet mixing terms.
The Lagrangian in the scalar sector is
\begin{eqnarray}\nonumber
\mathcal{L}&=&-\frac{1}{2}\partial_\mu S\partial^\mu S-\frac{1}{2}m_S^2 S^2
-D_\mu H^\dagger D^\mu H
\\ \nonumber
&&-\frac{\eta}{2}S^2\left(H^\dagger H-\frac{v_h^2}{2}\right)
\\ \label{eq:defL}
&&-\frac{\lambda}{4}\left(H^\dagger H-\frac{v_h^2}{2}\right)^2.
\end{eqnarray}
The derivatives $D_\mu$ are the appropriate $SU(2)\times U(1)$ covariant 
derivatives acting on the Higgs field. We assume perturbative 
singlet-Higgs coupling for our calculations of singlet annihilation cross 
sections in the non-relativistic limit. We report results in particular for
$\eta^2=0.1$ and for $\eta^2=0.01$. 
The singlet sector may also include a $Z_2$ symmetric singlet 
self-interaction $\sim\lambda_S S^4$ if the positive coupling $\lambda_S$ 
is weak enough such that its loop contributions can be neglected in the 
present perturbative calculation of singlet annihilation cross sections. 
The assumption of perturbative couplings in the non-relativistic limit is 
compatible with the fact that the $\beta$ functions are positive
in leading order in the couplings \cite{mura}.
We assume that the singlet vacuum expectation value vanishes, 
$v_s=\langle S\rangle=0$. Otherwise the singlet-Higgs coupling would 
yield a singlet-Higgs mixing term $\sim\eta v_s v_h sh$.

Later on we will also allow for a set of $N$ singlet states with
a global $O(N)$ symmetry and vanishing vacuum expectation values.
This ensures mass degeneracy and universality of the 
singlet-Higgs coupling strength $\eta$.

The Higgs-channel between dark matter and the standard model
was denoted as a Higgs portal in \cite{frank}. The model
provides a minimal renormalizable dark matter model\footnote{Other
recently proposed classes of minimal dark matter models introduce
heavy electroweak multiplets with a lightest neutral component
\cite{cirelliA,cirelli}, or another scalar messenger between
dark matter and standard matter \cite{mcd2}.}  
\cite{zee,mcd,cliff,mura,frank,zee2,maxim,hitoshi}. It has also been 
discussed as a model for quintessence in \cite{BR}. McDonald introduced 
the variant with a complex singlet \cite{mcd}, and effects of gauging 
the ensuing hidden U(1) symmetry are discussed in \cite{CNW}.
The presence of the singlet coupling obviously modifies the
Higgs effective potential and impacts electroweak symmetry breaking
\cite{hermann}, eventually triggering a strong first order phase
transition \cite{EQ,PRS}. 

A light electroweak singlet could reveal
itself as missing energy in collider based Higgs search experiments 
\cite{DR,dav1,SW,CNW} or in B decays \cite{maxim}.
March-Russell et al. pointed out that electroweak singlets and their
fermionic partners in supersymmetric theories can also be very heavy, 
with masses up to 30 TeV \cite{john}.

In the present paper we will be concerned with the prospects of 
observation of intermediate mass electroweak singlets through
their annihilation products in cosmic rays. We will take the
proposals of an electroweak singlet coupling through the Higgs portal
as a minimal renormalizable dark matter model seriously, and discuss
possible annihilation signals under the assumption $m_S\sim m_h$. 
The relevant mechanism for a sizable signal for an intermediate mass 
singlet would be annihilation through an intermediate Higgs boson into 
$W$ and $Z$ bosons. Because of the importance of the opening of the $W$ 
channel at 80 GeV, we will denote the mass range 
$80\,\mathrm{GeV}<m_S<1\,\mathrm{TeV}$ as the intermediate mass range 
for electroweak singlets. Proximity of $m_S$ to the $W$, $Z$, and Higgs
peaks in the annihilation cross section increases the
product $v\sigma$ substantially, thus potentially yielding a strongly enhanced
flux of annihilation products from the galactic halo even without invoking
boost factors from strong local dark matter overdensities. However, we will
see that the effect of the enhanced annihilation cross section is partially
compensated for in standard Lee-Weinberg theory for the creation of thermal
relics, because the requirement $\varrho_S=\varrho_{dm}$ together with
$m_S\sim m_h$ will require $N$-plets of electroweak singlet states, and
the net effect is a scaling of the flux $j\propto v\sigma/N$.

The strength of a direct galactic dark matter annihilation signal depends 
on the dark matter distribution in our galactic halo. The cosmic ray flux 
from a Navarro-Frenk-White (NFW) dark matter halo and a cored isothermal 
halo will be reviewed in section \ref{sec:halo}. Model independent limits 
on the flux will be discussed in \ref{sec:unit}.

Annihilation cross sections which follow from the singlet-Higgs coupling 
(\ref{eq:coupl1}) are reported in section \ref{sec:sigma}. 
The application of Lee-Weinberg theory for intermediate mass
electroweak singlets is discussed in section \ref{sec:LWabundance}.
Section \ref{sec:conc} summarizes our conclusions.

\section{The flux from the galactic halo}\label{sec:halo}

Annihilation of dark matter particles of mass $m_S$ and number 
density $n(\vec{r})$ generates a diffuse cosmic ray flux at our 
location $\vec{r}_\odot$
\cite{BDK}
\begin{eqnarray}\nonumber
j&=&\int d^3\vec{r}\frac{\nu n^2(\vec{r})}{4\pi|\vec{r}_{\odot}-\vec{r}|^2}
\\ \label{eq:flux1}
&&\times\frac{d\mathcal{N}(E,2m_S)}{dE}
\frac{\sigma v}{4\pi\,\mathrm{sr}}.
\end{eqnarray}
Here $\nu=1/2$ if the annihilating particles are Majorana to avoid
overcounting of collisions \cite{lars2}, and $\nu=1/4$ otherwise 
(assuming in the non-Majorana case equal amounts of dark matter and 
anti-matter in the halo). The case of interest to us is $\nu=1/2$.
The fragmentation function
\[
\mathcal{F}(E,E_{in})=\frac{d\mathcal{N}(E,E_{in})}{dE}
=\frac{1}{\sigma}\frac{d\sigma}{dE}
\]
gives the number of particles per energy interval and per annihilation
event for an event with initial energy $E_{in}$. 
Later on we will mostly focus on the photonic part 
$d\mathcal{N}_\gamma/dE=\sigma^{-1}d\sigma^{(\gamma)}/dE$, as charged and
hadronic components of dark matter annihilation products are masked by 
a relatively larger cosmic ray background than photons or neutrinos. 
However, in the present section we will also discuss the fraction of the 
total flux of cosmic rays from dark matter annihilation products compared 
to the cosmic rays flux $j_{CR}$ from all sources.

The factor $\sigma v$ in equation (\ref{eq:flux1}) is the product of 
annihilation cross section and relative speed in the non-relativistic 
limit. For processes at high redshift or annihilation of very low mass 
particles, thermal averaging $\langle\sigma v\mathcal{F}\rangle$ would 
have to be included. For high redshift sources, redshifting of 
$d\sigma/dE$ would also have to be included and the distance
$|\vec{r}_{\odot}-\vec{r}|$ has to be replaced by the luminosity 
distance. However, for annihilation of heavy particles in the galactic 
halo, equation (\ref{eq:flux1}) and its corresponding
line-of-sight counterparts below are perfectly adequate.

Equation (\ref{eq:flux1}) yields a cosmic ray flux averaged over all 
directions. If the detector is only sensitive to cosmic rays from a 
small solid angle $\Delta\Omega\ll 1$ sr, or if corresponding cuts can 
be applied, the observed flux per unit of solid angle is 
\cite{bergstrom} (see also \cite{sarkar}),
\begin{eqnarray}\nonumber
j_{\Delta\Omega}&=&
\int_0^\infty dx\int_{\Delta\Omega}d\vartheta d\varphi\sin\vartheta
\frac{\nu\tilde{n}^2(x,\vartheta,\varphi)}{4\pi}
\\ \label{eq:flux2}
&&\times\frac{d\mathcal{N}(E,2m_S)}{dE}
\frac{\sigma v}{\Delta\Omega}.
\end{eqnarray}
The vector $\vec{x}$ with length $x$ and direction $\vartheta,\varphi$
is related to the vector $\vec{r}$ in (\ref{eq:flux1}) through
$\vec{x}=\vec{r}-\vec{r}_\odot$, $\tilde{n}(\vec{x})\equiv n(\vec{r})$. 

The averaged flux (\ref{eq:flux1}) is recovered from equation 
(\ref{eq:flux2}) in the following way. The observed flux for aperture 
$\Delta\Omega\to 0$ is the integral along the line of sight 
$(\vartheta,\varphi)$,
\begin{eqnarray}\nonumber
j_{\Delta\Omega\to 0}(\vartheta,\varphi)&=&
\int_0^\infty dx\frac{\nu\tilde{n}^2(x,\vartheta,\varphi)}{4\pi}
\\ \label{eq:flux3}
&&\times\frac{d\mathcal{N}(E,2m_S)}{dE}\sigma v.
\end{eqnarray}
Averaging $j_{\Delta\Omega\to 0}(\vartheta,\varphi)$ over all directions
yields the diffuse flux (\ref{eq:flux1}),
\begin{eqnarray}\nonumber
j&=&\langle j_{\Delta\Omega\to 0}(\vartheta,\varphi)\rangle
\\ \nonumber
&=&
\int_0^\infty dx\int_0^\pi d\vartheta\int_0^{2\pi} d\varphi\sin\vartheta
\frac{\nu\tilde{n}^2(x,\vartheta,\varphi)}{4\pi}
\\ \nonumber
&&\times\frac{d\mathcal{N}(E,2m_S)}{dE}
\frac{\sigma v}{4\pi\,\mathrm{sr}}
\\ \nonumber
&=&\int d^3\vec{x}\frac{\nu\tilde{n}^2(\vec{x})}{4\pi |\vec{x}|^2}
\\ \label{eq:flux4}
&&\times\frac{d\mathcal{N}(E,2m_S)}{dE}
\frac{\sigma v}{4\pi\,\mathrm{sr}}.
\end{eqnarray} 

We will use equation (\ref{eq:flux1}) to estimate the averaged cosmic 
ray flux from dark matter annihilation in the galactic halo and 
equation (\ref{eq:flux3}) to estimate the diffuse cosmic ray flux 
along a line of sight orthogonal to the galactic plane. 

We refer to the differential fluxes (\ref{eq:flux1},\ref{eq:flux2}) as 
inclusive fluxes. Stable final states include
photons, neutrinos, electrons, positrons, protons and anti-protons.
The differential photon fluxes are found by substituting
$d\mathcal{N}(E,2m_S)/dE\to d\mathcal{N}_{\gamma}(E,2m_S)/dE$
in equations (\ref{eq:flux1}-\ref{eq:flux3}).

The density profile is assumed as an NFW profile \cite{nfw},
\begin{equation}\label{eq:nfw1}
\rho(r)=\frac{M}{r(r+r_s)^2},
\end{equation}
with a mass parameter 
$M=4.85\times 10^{10}M_\odot=5.41\times 10^{64}\,\mathrm{TeV}/c^2$
and a scale radius $r_s=21.5$ kpc. These parameters correspond to
the fit by Klypin et al. to the galactic halo \cite{klypin}, 
see also \cite{boyarski}.

Equation (\ref{eq:flux1}) can be evaluated analytically for
an NFW profile (we use $\nu=1/2$ in the following),
\begin{eqnarray}\nonumber
j&=&\int_0^\infty dr\frac{r}{r_{\odot}}
\ln\left(\frac{r+r_{\odot}}{|r-r_{\odot}|}\right)
\frac{\rho^2(r)}{4m_S^2}
\\ \nonumber
&&\times\frac{d\mathcal{N}(E,2m_S)}{dE}
\frac{\sigma v}{4\pi\,\mathrm{sr}}
\\ \nonumber
&=&\frac{M^2}{4m_S^2}\frac{d\mathcal{N}(E,2m_S)}{dE}
\frac{\sigma v}{4\pi\,\mathrm{sr}}
\frac{1}{r_{\odot}r_s^4}
\\ \nonumber
&&\times\left[\frac{\pi^2}{6}+
\mbox{L}_2\left(\frac{r_s}{r_s+r_{\odot}}\right)
+\mbox{L}_2\left(\frac{r_s-r_{\odot}}{r_s}\right)
\right.
\\ \nonumber
&&+\frac{1}{2}\ln^2\left(\frac{r_s+r_{\odot}}{r_s}\right)
+\frac{4r_sr_{\odot}(2r_s^2-r_{\odot}^2)}{3(r_s^2-r_{\odot}^2)^2}
\\ \label{eq:flux1b}
&&
-\left. 2r_{\odot}r_s\frac{9r_s^4-8r_s^2r_{\odot}^2
+3r_{\odot}^4}{3(r_s^2-r_{\odot}^2)^3}
\ln\left(\frac{r_s}{r_{\odot}}\right)\right]
\end{eqnarray}
where $\mbox{L}_2$ denotes Euler's dilogarithmic
function\footnote{Note that $\mbox{L}_2(x)=f(1-x)$ in \cite{AS}.
Although of no practical relevance, 
we point out that the singularity at $r_s=r_\odot$
cancels between the last two terms in equation (\ref{eq:flux1b}).}
\cite{bmp,AS}.

Substitution of data for the galactic halo yields
\begin{eqnarray}\nonumber
j&=&1.95\times 10^{14}\times
\left(\frac{\mathrm{TeV}}{m_S}\right)^2
\\ \nonumber
&&\times\frac{d\mathcal{N}(E,2m_S)}{dE}
\frac{\sigma v}{\mathrm{cm}^5\,\mathrm{sr}}
\\ \nonumber
&=&1.95\times 10^{14}\times
\left(\frac{\mathrm{TeV}}{m_S}\right)^3
\\ \label{eq:flux1c}
&&\times\frac{d\mathcal{N}(x,2m_S)}{dx}
\frac{\sigma v}{
\mathrm{TeV}\,\mathrm{cm}^5\,\mathrm{sr}},
\end{eqnarray}
with the scaled energy variable $x=E/m_S$.

Looking only along a line of sight orthogonal to the galactic 
plane to minimize background effects will cost approximately
a factor 3 in observed flux from the galactic halo,
\begin{eqnarray}\nonumber
j_\perp&=&
6.20\times 10^{13}\times
\left(\frac{\mathrm{TeV}}{m_S}\right)^3
\\ \label{eq:jperp}
&&\times\frac{d\mathcal{N}(x,2m_S)}{dx}
\frac{\sigma v}{
\mathrm{TeV}\,\mathrm{cm}^5\,\mathrm{sr}}.
\end{eqnarray}

Cirelli et al. also compared their calculations with other 
density profiles \cite{cirelli}, using the assumption of same
local dark matter density. This yields for the isothermal cored
profile
\[
\varrho=\frac{\mu}{r^2 + r_c^2},\quad r_c=5\,\mathrm{kpc},
\]
a parameter
$\mu=6.92\times 10^{62}\,\mathrm{GeV}/\mathrm{kpc}
=2.24\times 10^{41}\,\mathrm{GeV}/\mathrm{cm}$.

The absence of the central cusp reduces the flux
averaged over all directions,
\begin{eqnarray}\nonumber
j_{isc}&=&\frac{\pi \mu^2}{8 m_S^2 r_c(r^2_c+r^2_\odot)}
\frac{d\mathcal{N}(E,2m_S)}{dE}
\frac{\sigma v}{4\pi\,\mathrm{sr}}
\\ \nonumber
&=&1.20\times 10^{14}\times
\left(\frac{\mathrm{TeV}}{m_S}\right)^3
\\ \label{eq:isc}
&&\times\frac{d\mathcal{N}(x,2m_S)}{dx}
\frac{\sigma v}{
\mathrm{TeV}\,\mathrm{cm}^5\,\mathrm{sr}},
\end{eqnarray}
but the flux along a line of sight perpendicular to the
galactic plane is virtually unchanged and slightly increased 
due to the weaker local gradient in the dark matter 
distribution,
\begin{eqnarray}\nonumber
j^{(isc)}_\perp&=&
6.36\times 10^{13}\times
\left(\frac{\mathrm{TeV}}{m_S}\right)^3
\\ \label{eq:iscperp}
&&\times\frac{d\mathcal{N}(x,2m_S)}{dx}
\frac{\sigma v}{
\mathrm{TeV}\,\mathrm{cm}^5\,\mathrm{sr}}.
\end{eqnarray}
We will report numerical results for the flux $j$ from
equation (\ref{eq:flux1c}). The other fluxes can then easily be 
derived from equations (\ref{eq:jperp}-\ref{eq:iscperp}).

The fragmentation function into all final states must 
satisfy the energy sum rule
\[
\int_0^1 dx x\frac{d\mathcal{N}(x)}{dx}=2.
\] 
An often used parametrization is
\begin{equation}\label{eq:mlla}
\frac{d\mathcal{N}(x)}{dx}=
\frac{2x^\alpha(1-x)^\beta}{B(\alpha+2,\beta+1)}.
\end{equation}
We will use fiducial values $\alpha=-1.5$, $\beta=2$
for numerical estimates.
The two slope behavior of $\log(d\mathcal{N}(x)/dx)$
versus $x$ (see e.g. Fig. 17.1 in the review \cite{biebel})
also suggests a phenomenological fit
\begin{equation}\label{eq:hrs}
\frac{d\mathcal{N}(x)}{dx}=
A\exp(-\alpha x)+B\exp(-\beta x).
\end{equation}
The values found by the HRS Collaboration 
in $e^+ e^-$ annihilation at $\sqrt{s}=29$ GeV
correspond to normalized values $A=360.5$, $\alpha=28.4$,
$B=97.2$ and $\beta=7.91$ \cite{hrs}. A disadvantage
of these 2-temperature distributions is that they are
very small, but do not vanish at $x=1$. However, they
work very well between $0.1<x<0.9$.

The shape of fragmentation functions is only weakly
energy dependent between 12 GeV and 202 GeV \cite{biebel},
and dominated by the fragmentation properties of intermediate
partons. Therefore we also use (\ref{eq:hrs})
for numerical work besides (\ref{eq:mlla}).

We will use the differential photon fragmentation function 
proposed in \cite{lars1},
\begin{equation}\label{eq:lars1}
\frac{d\mathcal{N}_{\gamma}(x)}{dx}=
\frac{0.42\exp(-8x)}{x^{1.5}+0.00014},
\end{equation}
for the differential photon spectrum.
This corresponds to about 11\% photon energy yield.

\section{Model independent bounds on the cosmic ray flux from
dark matter annihilation}\label{sec:unit}

Although we are primarily interested in cosmic ray fluxes from
the Higgs portal coupling (\ref{eq:coupl1}), we would also like
to point out that model independent estimates for cosmic rays
from dark matter annihilation arise from unitarity limits and from 
limits on neutrino fluxes. Another model independent limit arises
from halo stability \cite{KKT}. However, this limit has been 
superseded by the neutrino limit for dark matter masses heavier than 
0.1 GeV \cite{BBM,yuksel}.

The unitarity limit on s wave annihilation cross sections 
$\sigma\le 4\pi/k^2$ \cite{steve2,unit} implies
\begin{eqnarray}\nonumber
\sigma v&\le&4.40\times 10^{-19}\frac{\mathrm{cm}^3}{\mathrm{s}}
\\ \label{eq:limit1}
&&\times\left(\frac{\mathrm{TeV}}{m_S}\right)^2
\frac{100\,\mathrm{km}/\mathrm{s}}{v}.
\end{eqnarray}
Substitution in equation (\ref{eq:flux1c}) yields a limit on the 
diffuse cosmic ray flux from galactic dark matter annihilation
\begin{eqnarray}\nonumber
j_S&\le&\frac{8.57\times 10^{-5}}{
\mathrm{TeV}\,\mathrm{cm}^2\,\mathrm{s}\,\mathrm{sr}}
\frac{d\mathcal{N}(x)}{dx}
\\ \label{eq:limit2}
&&\times\left(\frac{\mathrm{TeV}}{m_S}\right)^5
\frac{100\,\mathrm{km}/\mathrm{s}}{v}.
\end{eqnarray}

Beacom et al. and Y\"uksel et al. recently found that limits on the 
diffuse cosmic neutrino signal and the halo signal can be 
used to impose stronger limits on the dark matter annihilation 
cross section for dark matter masses below 10 TeV \cite{BBM,yuksel}.
Between 10 GeV and 10 TeV, the cosmic and the isotropic
halo neutrino flux limits approximately reduce the 
limit on the annihilation cross section $\sim m^2$. 
We take this into account through a correction factor
\[
\beta_\nu=\left\{\begin{array}{ll}
\left(\frac{m_S}{10\,\mathrm{TeV}}\right)^2, & 
10\,\mathrm{GeV}\le m_S\le 10\,\mathrm{TeV},\\
1, & m_S> 10\,\mathrm{TeV}.\\
\end{array}
\right.
\]

Comparison with the cosmic ray flux below 1 PeV \cite{hegra}
\begin{eqnarray*}
j_{CR}&=&\frac{2.582\times 10^{-5}}
{\mathrm{TeV}\,\mathrm{cm}^2\,\mathrm{s}\,\mathrm{sr}}
\left(\frac{\mathrm{TeV}}{E}\right)^{2.68}
\\
&=&\frac{1.236\times 10^{-5}}
{\mathrm{GeV}\,\mathrm{cm}^2\,\mathrm{s}\,\mathrm{sr}}
\left(\frac{100\,\mathrm{GeV}}{E}\right)^{2.68}
\end{eqnarray*}
shows that a galactic dark matter annihilation signal could reach 
several per cent of the total differential cosmic ray flux if the 
annihilation cross section could get close to the upper limits either 
through Sommerfeld enhancement or through resonance effects. We find
\begin{eqnarray*}
\frac{j_S}{j_{CR}}&\le& \beta_\nu\times
3.32 x^{2.68}\frac{d\mathcal{N}(x)}{dx}
\\
&&\times\left(\frac{\mathrm{TeV}}{m_S}\right)^{2.32}
\frac{100\,\mathrm{km}/\mathrm{s}}{v},
\end{eqnarray*}
e.g. for $m_S=100$ GeV,
\[
\frac{j_S}{j_{CR}}\le
6.94\times 10^{-2} x^{2.68}\frac{d\mathcal{N}(x)}{dx}\times
\frac{100\,\mathrm{km}/\mathrm{s}}{v}.
\]

The maximum of $x^{2.68}d\mathcal{N}(x)/dx$ for the  fragmentation
function (\ref{eq:mlla}) is 0.23 for $x=0.37$.  For the fragmentation 
function (\ref{eq:hrs}) the maximum of $x^{2.68}d\mathcal{N}(x)/dx$ is 
0.37 for $x=0.34$.

However, cosmic rays are strongly dominated by charged particles,
while dark matter annihilation products are expected to contain 
a relatively higher neutral component of photons and neutrinos.
$\gamma$ ray and neutrino observatories are therefore the primary
search tools for dark matter annihilation products. We will use 
the diffuse photon background published by EGRET between 30 MeV 
and 120 GeV \cite{egret},
\begin{eqnarray}\nonumber
j_{\gamma,E}&=&\frac{6.89\times 10^{-10}}
{\mathrm{TeV}\,\mathrm{cm}^2\,\mathrm{s}\,\mathrm{sr}}
\left(\frac{\mathrm{TeV}}{E}\right)^{2.10}
\\ \label{eq:egret}
&=&\frac{8.68\times 10^{-11}}
{\mathrm{GeV}\,\mathrm{cm}^2\,\mathrm{s}\,\mathrm{sr}}
\left(\frac{100\,\mathrm{GeV}}{E}\right)^{2.10},
\end{eqnarray}
for benchmarking. This is the diffuse photon flux observed
by EGRET after subtraction of conventional galactic sources.
It has been pointed out that improved models for standard
sources could reduce the EGRET signal, see 
\cite{disc1} and references there. The sensitivity calibration
beyond 1 GeV has also been called into question 
\cite{disc2}. The energy range between 1 GeV and 100 GeV is
particularly relevant for continuous photon signals from dark 
matter in the mass range of interest here, 
$80\,\mathrm{GeV}<m_S<1\,\mathrm{TeV}$, and we will see that signal
levels may be small. Reliable further subtractions of standard
sources from the diffuse $\gamma$ ray background signal, and a lower
total signal, would help to identify or constrain a possible dark 
matter signal. EGRET sets a useful benchmark until GLAST/LAT publishes 
data on the diffuse $\gamma$ ray background.

Normalizing the averaged photon flux following from equation
(\ref{eq:flux1c}) to the EGRET flux (\ref{eq:egret}) yields
\begin{eqnarray*}
\frac{j_\gamma}{j_{\gamma,E}}&=&
2.83\times 10^{23} x^{2.1}\frac{d\mathcal{N}_\gamma(x)}{dx}
\\
&&\times\left(\frac{\mathrm{TeV}}{m_S}\right)^{0.9}
\frac{\sigma v}{\mathrm{cm}^3\,\mathrm{s}^{-1}}.
\end{eqnarray*}
The maximum of $x^{2.1}d\mathcal{N}_\gamma(x)/dx$ is $0.05$
near $x\sim 0.08$. Therefore any annihilation cross section
of a dark matter particle with a mass below 1.5 TeV is 
constrained to
\begin{equation}\label{eq:condEG}
\sigma v<10^{-23}\frac{\mathrm{cm}^3}{\mathrm{s}}
\times\left(\frac{m_S}{100\,\mathrm{GeV}}\right)^{0.9}.
\end{equation}

\section{Annihilation cross sections for the 
electroweak singlet}\label{sec:sigma}

The coupling (\ref{eq:coupl1}) reduces in unitary gauge to
\begin{equation}\label{eq:coupl2}
H_{Sh}=\frac{\eta v_h}{2}\int d^3\vec{x} S^2 h
+\frac{\eta}{4}\int d^3\vec{x} S^2 h^2.
\end{equation}
 For the annihilation of the electroweak singlet through an 
intermediate Higgs, we also need the couplings
\begin{equation}\label{eq:couplhh}
H_{hh}=\int d^3\vec{x}\frac{m_h^2}{2v_h}
\left(h^3+\frac{h^4}{4v_h}\right),
\end{equation}
\begin{equation}\label{eq:couplhf}
H_{fh}=\int d^3\vec{x}\sum_f\frac{m_f}{v_h}h\overline{f}
\cdot f
\end{equation}
and
\begin{eqnarray}\nonumber
H_{W,Zh}&=&\int d^3\vec{x}\left(
2\frac{m_W{}^2}{v_h{}^2} W^- W^+ +\frac{m_Z{}^2}{v_h{}^2} Z^2\right)
\\ \label{eq:couplhW}
&&\times\left(v_h h+\frac{h^2}{2}\right).
\end{eqnarray}

These couplings yield the following annihilation cross sections 
in the non-relativistic limit,
\begin{eqnarray*}
v\sigma_{SS\to hh}&=&
\eta^2\frac{\sqrt{m_{S}{}^2-m_h{}^2}}{16\pi m_{S}{}^3}
\\
&&\times
\left|\frac{2m_S{}^2 + m_h{}^2}{4m_S{}^2-m_h{}^2+\mathrm{i}m_h\Gamma_h}\right.
\\
&&
-\left.
\frac{2\eta v_h{}^2}{2m_S{}^2-m_h{}^2}\right|^2,
\end{eqnarray*}
\begin{eqnarray*}
v\sigma_{SS\to f\overline{f}}
&=&\eta^2\frac{N_c m_f{}^2}{
4\pi m_S{}^3}
\\
&&\times
\frac{\sqrt{m_S{}^2-m_f{}^2}^3}{
(4m_S{}^2-m_h{}^2)^2+m_h{}^2\Gamma_h{}^2},
\end{eqnarray*}
\begin{eqnarray*}
v\sigma_{SS\to WW}
&=&\eta^2\frac{\sqrt{m_S{}^2-m_{W}{}^2}}{4\pi m_S{}^3}
\\
&&\times
\frac{3m_{W}{}^4-4m_{W}{}^2m_S{}^2+4m_S{}^4}{
(4m_S{}^2-m_h{}^2)^2+m_h{}^2\Gamma_h{}^2},
\end{eqnarray*}
\begin{eqnarray*}
v\sigma_{SS\to ZZ}
&=&\eta^2\frac{\sqrt{m_S{}^2-m_{Z}{}^2}}{8\pi m_S{}^3}
\\
&&\times\frac{3m_{Z}{}^4-4m_{Z}{}^2m_S{}^2+4m_S{}^4}{
(4m_S{}^2-m_h{}^2)^2+m_h{}^2\Gamma_h{}^2},
\end{eqnarray*}
with $N_c=3$ for quarks and $N_c=1$ for leptons. The 
$\sigma_{SS\to f\overline{f}}$ cross section is summed over final spin 
states. The cross section $\sigma_{SS\to hh}$ contains the scattering 
amplitude from the $S^2h^2$ contact vertex in equation (\ref{eq:coupl2}), 
the scattering amplitude from the $s$-channel contribution with an
intermediate Higgs boson from the $S^2h$ vertex in equation 
(\ref{eq:coupl2}) and the $h^3$ vertex in equation (\ref{eq:couplhh}),
and the $t$-channel and $u$-channel amplitudes with an intermediate singlet 
from the $S^2h$ vertex. Those amplitudes are with the normalization
$S_{fi}=\delta_{fi}-\mathrm{i}\mathcal{M}_{fi}\delta^4(p_1+p_2-k_1-k_2)$,
\begin{eqnarray*}
\mathcal{M}_{SS\to hh}^{(1)}&=&
\frac{\eta}{16\pi^2}
\\
&&\times\frac{1}{\sqrt{E_S(\vec{k}_1)E_S(\vec{k}_2)
E_h(\vec{p}_1)E_h(\vec{p}_2)}},
\\
\mathcal{M}_{SS\to hh}^{(2)}&=&
-\frac{3\eta m_h{}^2}{16\pi^2}
\\
&&\times\frac{1}{\sqrt{E_S(\vec{k}_1)E_S(\vec{k}_2)
E_h(\vec{p}_1)E_h(\vec{p}_2)}}\\
&&\times\frac{1}{(k_1+k_2)^2+m_h{}^2-\mathrm{i}\epsilon},
\\
\mathcal{M}_{SS\to hh}^{(3)}&=&
-\frac{\eta^2 v_h{}^2}{8\pi^2}
\\
&&\times\frac{1}{\sqrt{E_S(\vec{k}_1)E_S(\vec{k}_2)
E_h(\vec{p}_1)E_h(\vec{p}_2)}}\\
&&\times\left(\frac{1}{(k_1-p_1)^2+m_S{}^2-\mathrm{i}\epsilon}\right.
\\
&&+\left.\frac{1}{(k_1-p_2)^2+m_S{}^2-\mathrm{i}\epsilon}\right).
\end{eqnarray*}

Note that due to the constraint $m_S{}^2>m_h{}^2$ for $SS\to hh$, 
neither the $s$-channel resonance for $m_S=m_h/2$ nor the $t$-channel 
or $u$-channel resonance for $m_S=m_h/\sqrt{2}$ can be realized for 
$\sigma_{SS\to hh}$. Furthermore, $\Gamma_S=0$ in our models. 

The amplitude for annihilation of two singlets with momenta $\vec{k}_1$
and $\vec{k}_2$ into massive vector bosons with momenta $\vec{p}_1$,
$\vec{p}_2$ and polarizations $\epsilon^{(\alpha)}(\vec{p}_1)$,
$\epsilon^{(\beta)}(\vec{p}_2)$ is proportional to $m_W{}^2$,
\begin{eqnarray} \nonumber
&&\mathcal{M}_{SS\to WW}=-\frac{\eta m_W{}^2}{8\pi^2}
\\ \nonumber
&&\times\frac{1}{
\sqrt{E_S(\vec{k}_1)E_S(\vec{k}_2)E_W(\vec{p}_1)E_W(\vec{p}_2)}}
\\ \label{eq:MSStoWW}
&&\times
\frac{\epsilon^{(\alpha)}(\vec{p}_1)\cdot\epsilon^{(\beta)}(\vec{p}_2)}{
(k_1+k_2)^2+m_h{}^2-\mathrm{i}\varepsilon}
\end{eqnarray}
However, the longitudinal polarization vector for massive vector bosons
comes with a factor $m_W{}^{-1}$, e.g. the polarization vectors for 
$\vec{p}$ in $z$-direction are
\[
\epsilon^{(1)}(\vec{p})=(0,1,0,0),\quad
\epsilon^{(2)}(\vec{p})=(0,0,1,0),
\]
\[
\epsilon^{(3)}(\vec{p})=
\frac{1}{m_{W}}\left(|\vec{p}|,0,0,\sqrt{\vec{p}^2+m_{W}{}^2}\right).
\]
The tensor product of massive polarization vectors
\[
\sum_\alpha\epsilon^{(\alpha)\mu}(\vec{p})
\otimes\epsilon^{(\alpha)\nu}(\vec{p})
=\eta^{\mu\nu}+\frac{p^\mu p^\nu}{m_{W}{}^2}
\]
implies
\begin{eqnarray*}
&&\sum_{\alpha,\beta}\left(\epsilon^{(\alpha)}(\vec{p}_1)
\cdot\epsilon^{(\beta)}(\vec{p}_2)\right)^2
\\
&&=\left(\eta^{\mu\nu}+\frac{p_1^\mu p_1^\nu}{m_{W}{}^2}\right)
\left(\eta_{\nu\mu}+\frac{p_{2\nu} p_{2\mu}}{m_{W}{}^2}\right)
\\
&&=2+\frac{(p_1\cdot p_2)^2}{m_{W}{}^4}.
\end{eqnarray*}
Together with the amplitude (\ref{eq:MSStoWW}) this yields the cross
section $\sigma_{SS\to WW}$. An equivalent way to understand why the 
cross sections $\sigma_{SS\to WW}$ and $\sigma_{SS\to ZZ}$ do not 
vanish in the limit of vanishing coupling 
$v_h{}^2\sim m_{W}{}^2\sim m_{Z}{}^2\to 0$ are the residual Goldstone 
boson modes $h^\pm$ and $\chi^0$ from the Higgs field, which would 
contribute to singlet annihilation in the $SU(2)\times U(1)$ symmetric 
limit.

Standard model decay widths of the Higgs particle are small
unless the Higgs boson is heavy enough to decay into weak
gauge bosons \cite{DGH}, e.g. 
$\Gamma_h(m_h=115\,\mathrm{GeV})=4.56\,\mathrm{MeV}$,
$\Gamma_h(m_h=160\,\mathrm{GeV})=6.37\,\mathrm{MeV}$, 
$\Gamma_h(m_h=165\,\mathrm{GeV})=247\,\mathrm{MeV}$, 
$\Gamma_h(m_h=203\,\mathrm{GeV})=1.54\,\mathrm{GeV}$. 

The energy dependent cross sections $\sigma v(K)$ in order
$\mathcal{O}(\eta^2)$ in all cases correspond to the substitution
\[
m_S^2\to (K+m_S)^2
\]
in the equations for $v\sigma$, where $K$ is the kinetic
energy of an electroweak singlet. The non-relativistic limits
for $v\sigma$ are therefore not only excellent approximations
for $T\ll m_S$, but also provide upper limits on the thermal 
averages at high temperature
\[
\langle\sigma v\rangle
=\frac{\int_0^\infty dK\frac{(K+m_S)\sqrt{K(K+2m_S)}}{
\exp[(K+m_S)/T]-1}\sigma v(K)}{
\int_0^\infty dK\frac{(K+m_S)\sqrt{K(K+2m_S)}}{
\exp[(K+m_S)/T]-1}}.
\]
The temperature dependence is very weak. The $\mathcal{O}(\eta^2)$ 
cross section for $m_S=200$ GeV is only reduced from
$\sigma v=1.04\eta^2\times 10^{-23}\,\mathrm{cm}^3/\mathrm{s}$
at $T=0$ to $\langle\sigma v\rangle
=0.39\eta^2\times 10^{-23}\,\mathrm{cm}^3/\mathrm{s}$ at $T=100$ GeV.

The tree-level cross sections reported here yield cross sections of 
order $\sigma v\sim\eta^2\times 10^{-23}\,\mathrm{cm}^3/\mathrm{s}$
in the mass range $m_S\sim m_h$, see also figures 
\ref{fig:vsigma1}-\ref{fig:vsheavy01} below. In addition there are also 
loop suppressed photon lines at 
$E_\gamma=m_S$ from a $\gamma\gamma$ final state, and at 
$E_\gamma=m_S-(m_Z{}^2/4m_S)$ from a $\gamma Z$ state. These 
contributions are of order of a per cent compared to the continuous 
spectrum, because the loop amplitudes contain two vertices of
electroweak strength or the direct $W^+ W^-\to\gamma\gamma$ contact 
vertex. The contribution to $v\sigma_{SS\to\gamma\gamma}$ through a 
$W^+ W^-$ loop with contact vertex e.g. is approximately of order
(neglecting logarithmic mass dependencies)
\begin{eqnarray*}
v\sigma_{SS\to\gamma\gamma}&\sim&\frac{32\pi\alpha^2\eta^2 m_W{}^4}{
m_S{}^{2}[(4m_S{}^2-m_h{}^2)^2+m_h{}^2\Gamma_h{}^2]}
\\
&\sim& 10^{-3}v\sigma_{SS}.
\end{eqnarray*}

The continuous spectrum yields a conservatively estimated photon energy 
yield of order 10\% (see equation (\ref{eq:lars1})). Together with the 
ratio of cross sections, this implies that in terms of photon energy 
yields from singlet annihilation, the line spectrum should be of order 
of a few per cent of the continuous spectrum. Therefore we focus on the 
continuous spectrum in the present investigation. However, further study 
of the loop induced line spectrum is also of interest. 

\subsection{Annihilation signal from a heavy electroweak 
singlet in tree level approximation}\label{sec:fluxheavy}

If we assume a Higgs mass limit $m_h\le 203$ GeV
from electroweak analysis \cite{PDG} and absence
of a very heavy fourth generation below $m_S$, we
get in leading order the following 
annihilation cross section for a heavy ($m_S>1$ TeV) 
electroweak singlet coupling through the Higgs portal,
\begin{eqnarray}\nonumber
v\sigma_{SS}&\simeq&\frac{7\eta^2}{64\pi m_S^2}
\\ \label{eq:anniSS}
&=&4.06\eta^2\times 10^{-25}
\left(\frac{\mathrm{TeV}}{m_S}\right)^2
\frac{\mathrm{cm}^3}{\mathrm{s}}.
\end{eqnarray}


This translates into a cosmic ray flux from
heavy electroweak singlet annihilation 
through the Higgs portal at tree level,
\[
j_S\simeq\frac{7.92\eta^2\times 10^{-11}}{
\mathrm{TeV}\,\mathrm{cm}^2\,\mathrm{s}\,\mathrm{sr}}
\left(\frac{\mathrm{TeV}}{m_S}\right)^5
\frac{d\mathcal{N}(x)}{dx}.
\]

This is very small compared to the galactic background flux 
from supernovae and their remnants, and again photon signals
are expected to be more sensitive. Normalizing to the
EGRET flux formally yields
\[
\frac{j_\gamma}{j_{\gamma,E}}
=0.115\eta^2 x^{2.1}\frac{d\mathcal{N}_\gamma(x)}{dx}
\left(\frac{\mathrm{TeV}}{m_S}\right)^{2.9}.
\]
Due to the energy limit of EGRET this should only be used 
for $x<0.12\,\mathrm{TeV}/m_S$. For $m_S=1$ TeV and
$E\simeq 76$ GeV we find
$j_\gamma/j_{\gamma,E}\simeq 5.6\eta^2\times 10^{-3}$.

 For heavier singlets, March-Russell et al. have pointed out 
that Sommerfeld enhancement due to scalar exchange between 
annihilating supersymmetric singlets can boost annihilation cross 
sections by factors $10^3$ to $10^5$ for singlet masses between 
1 TeV and 30 TeV \cite{john}. Sommerfeld enhanced 
annihilation signals from heavy electroweak 
multiplets have recently been discussed by Cirelli et al.
\cite{cirelli}.
In the present paper, we will instead focus on electroweak singlet 
annihilation in the intermediate mass range
 $80\,\mathrm{GeV}< m_S<1\,\mathrm{TeV}$, when $\sigma_{SS}$ is enhanced 
due to proximity to the $W$, $Z$ and Higgs peaks in the cross section.

\subsection{Enhancement of electroweak singlet annihilation 
for $m_S\sim m_h$ due to the $W$, $Z$ and Higgs peaks}\label{sec:peaks}

Electroweak singlet annihilation through an intermediate Higgs can be 
strongly enhanced when the channels $SS\to WW$ and $SS\to ZZ$ open up. 
This is especially relevant when $m_S\sim m_h$, which is the primary 
mass range of interest in our present investigation.

We will primarily use Higgs mass values $m_h=115$ GeV from the direct 
search limit \cite{PDG}, and $m_h=160$ GeV in the range of highest 
sensitivity for search for a light Standard Model Higgs boson at the 
Tevatron \cite{teva}. The value $m_h=160$ GeV is also in the preferred 
mass range for minimal dark matter models identified in reference 
\cite{mura}.

Compared to $\sigma_{SS\to WW}$ the contributions of light quarks and 
leptons to $\sigma_{SS}$ are of order $10^{-4}$ and the contributions of 
heavy quarks and leptons are of order $10^{-2}$ in the mass range of 
interest. We will include c, b and t quarks with masses $m_c=1.25$ GeV, 
$m_b=4.20$ GeV and $m_t=172.5$ GeV in the calculation of $v\sigma_{SS}$.
The $\tau$ lepton will be included with $m_\tau=1.78$ GeV.

\begin{figure}[htb]
\begin{center}
\scalebox{0.38}{\includegraphics{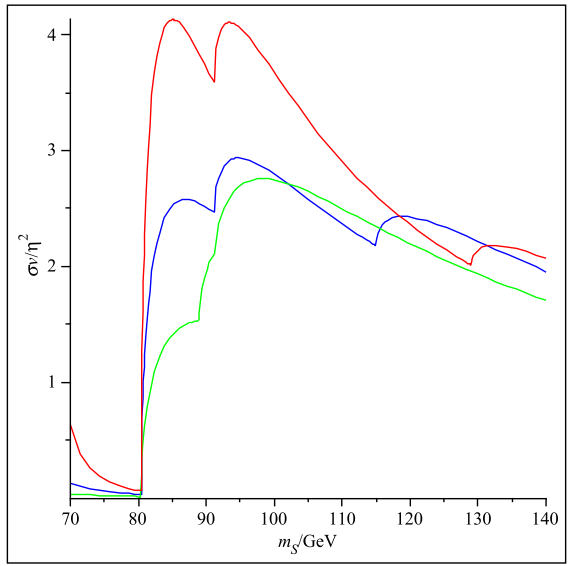}}
\end{center}
\caption{\it The cross section $\eta^{-2}v\sigma_{SS}$ in units
of $10^{-23}\,\mathrm{cm}^3/\mathrm{s}$ for very weak coupling
$\eta^2\lesssim 0.01$. The red (initially and finally
highest) curve is for $m_h=129\,\mathrm{GeV}$, the blue (initially and 
finally middle) curve is for $m_h=115\,\mathrm{GeV}$, and the green 
(initially and finally lowest) curve is for $m_h=89\,\mathrm{GeV}$.}
\label{fig:vsigma1}     
\end{figure}

\begin{figure}[htb]
\begin{center}
\scalebox{0.38}{\includegraphics{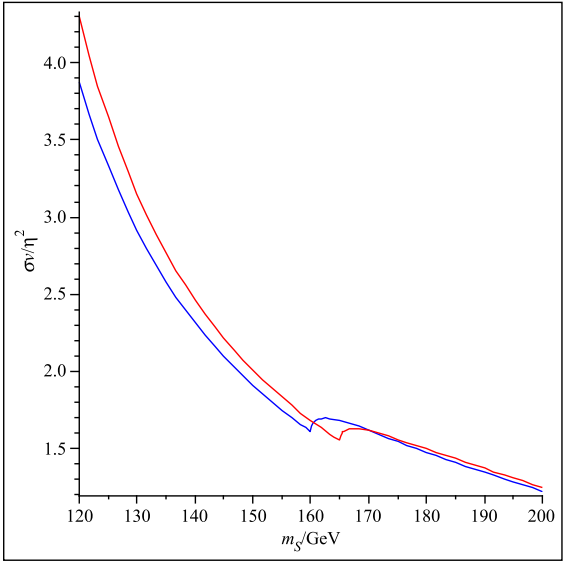}}
\end{center}
\caption{\it The cross section $\eta^{-2}v\sigma_{SS}$ in units
of $10^{-23}\,\mathrm{cm}^3/\mathrm{s}$ for very weak coupling
$\eta^2\lesssim 0.01$. 
The blue (initially and finally lower) curve is the cross section
for $m_h=160\,\mathrm{GeV}$.
The red (initially and finally upper) curve is the cross section
for $m_h=165\,\mathrm{GeV}$.}
\label{fig:vsigma3}     
\end{figure}

The effect of the $t$-channel plus $u$-channel amplitude 
$\mathcal{M}_{SS\to hh}^{(3)}$ is small for very weak coupling
$\eta^2\lesssim 0.01$. The $\mathcal{O}(\eta^2)$ cross sections
for very weak coupling are displayed for several Higgs mass
values in figures \ref{fig:vsigma1} and \ref{fig:vsigma3}.

 Figure \ref{fig:vsigma1} shows the scaled cross section 
$\eta^{-2}v\sigma_{SS}$ for $m_h=115$ GeV and
$70\,\mathrm{GeV}<m_S<140\,\mathrm{GeV}$. The cross sections for 
$m_h=89$ GeV and $m_h=129$ GeV are included for comparison.
The edge at 80 GeV arises from the opening of the
$W$ channel $SS\to WW$, the edge at 91 GeV arises from
the $Z$ channel, and the edge at $m_S=m_h$
arises from the Higgs channel $SS\to hh$.

The scaled cross sections $\eta^{-2}v\sigma_{SS}$ 
for $m_h=160$ GeV and $m_h=165$ GeV are displayed in figure 
\ref{fig:vsigma3}.

\begin{figure}[htb]
\begin{center}
\scalebox{0.38}{\includegraphics{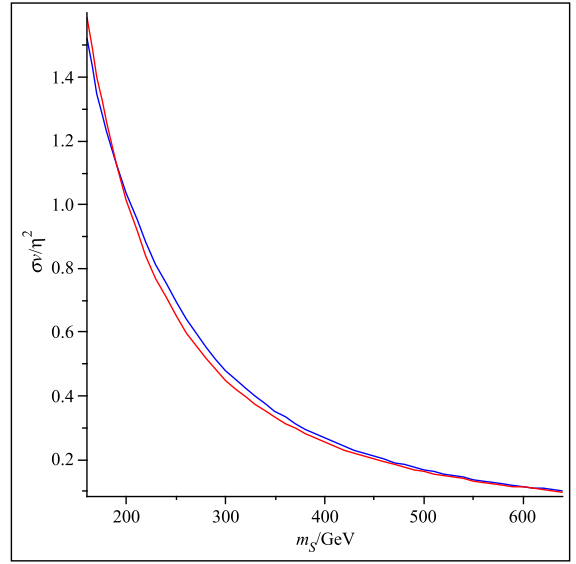}}
\end{center}
\caption{\it The cross section $\eta^{-2}v\sigma_{SS}$ in units
of $10^{-23}\,\mathrm{cm}^3/\mathrm{s}$ in the mass range
$160\,\mathrm{GeV}<m_S<640\,\mathrm{GeV}$ for very weak coupling
$\eta^2\lesssim 0.01$.
The red (initially upper) curve is the asymptotic cross section
(\ref{eq:anniSS}). The blue (initially lower) curve is for 
$m_h=115\,\mathrm{GeV}$. The cross section for $m_h=160\,\mathrm{GeV}$ 
also differs by less than 10\% from the aymptotic limit for 
$m_S>400\,\mathrm{GeV}$.}
\label{fig:vsigma2}     
\end{figure}

The cross sections for larger values of $m_S$ approach the asymptotic limit
(\ref{eq:anniSS}), see figure \ref{fig:vsigma2}.

\begin{figure}[htb]
\begin{center}
\scalebox{0.38}{\includegraphics{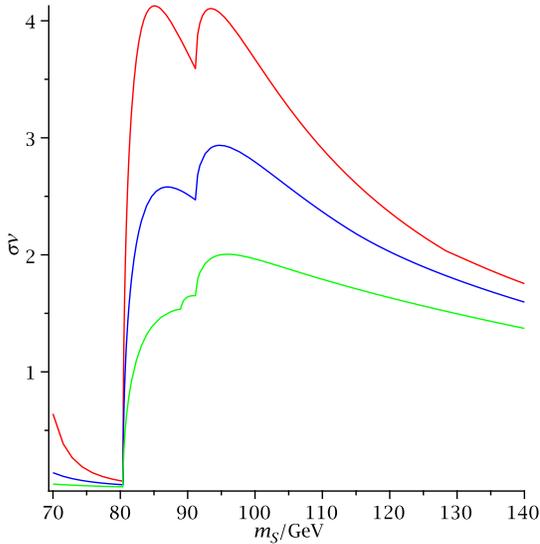}}
\end{center}
\caption{\it The cross section $v\sigma_{SS}$ in units
of $10^{-24}\,\mathrm{cm}^3/\mathrm{s}$ for coupling
$\eta^2=0.1$. The red (initially and finally
highest) curve is for $m_h=129\,\mathrm{GeV}$, the blue (initially and 
finally middle) curve is for $m_h=115\,\mathrm{GeV}$, and the green 
(initially and finally lowest) curve is for $m_h=89\,\mathrm{GeV}$.}
\label{fig:vslight01}     
\end{figure}

\begin{figure}[htb]
\begin{center}
\scalebox{0.38}{\includegraphics{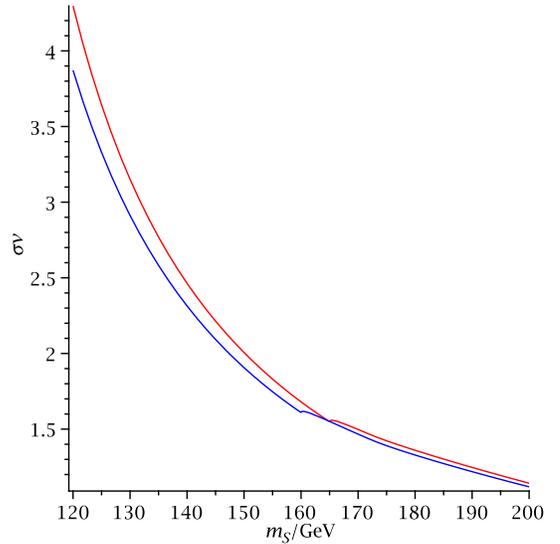}}
\end{center}
\caption{\it The cross section $v\sigma_{SS}$ in units
of $10^{-24}\,\mathrm{cm}^3/\mathrm{s}$ for coupling
$\eta^2=0.1$. 
The blue (initially and finally lower) curve is the cross section
for $m_h=160\,\mathrm{GeV}$.
The red (initially and finally upper) curve is the cross section
for $m_h=165\,\mathrm{GeV}$.}
\label{fig:vsheavy01}     
\end{figure}

 For electroweak strength coupling $\eta^2=0.1$, the $t$-channel plus 
$u$-channel amplitude $\mathcal{M}_{SS\to hh}^{(3)}$ suppresses the
Higgs threshold in the cross sections, see figures
\ref{fig:vslight01} and \ref{fig:vsheavy01}.

\begin{figure}[htb]
\begin{center}
\scalebox{0.38}{\includegraphics{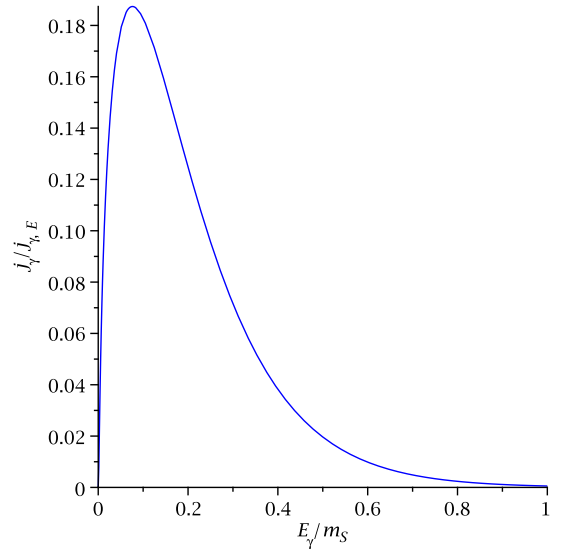}}
\end{center}
\caption{\it The photon flux $j_\gamma$ in the continuous spectrum for 
$m_h=115\,\mathrm{GeV}$, $m_S=120\,\mathrm{GeV}$, $\eta^2=0.1$ and 
$E_\gamma<120\,\mathrm{GeV}$ in units of the EGRET flux.}
\label{fig:phot115}     
\end{figure}

The continuous photon spectrum for $E_\gamma<120$ GeV in units of the EGRET flux 
is displayed in figure \ref{fig:phot115} for $m_S=120$ GeV, $m_h=115$ GeV,
$\eta^2=0.1$ and $v\sigma_{SS}=2.03\times 10^{-24}\,\mathrm{cm}^3/\mathrm{s}$.

\begin{figure}[htb]
\begin{center}
\scalebox{0.38}{\includegraphics{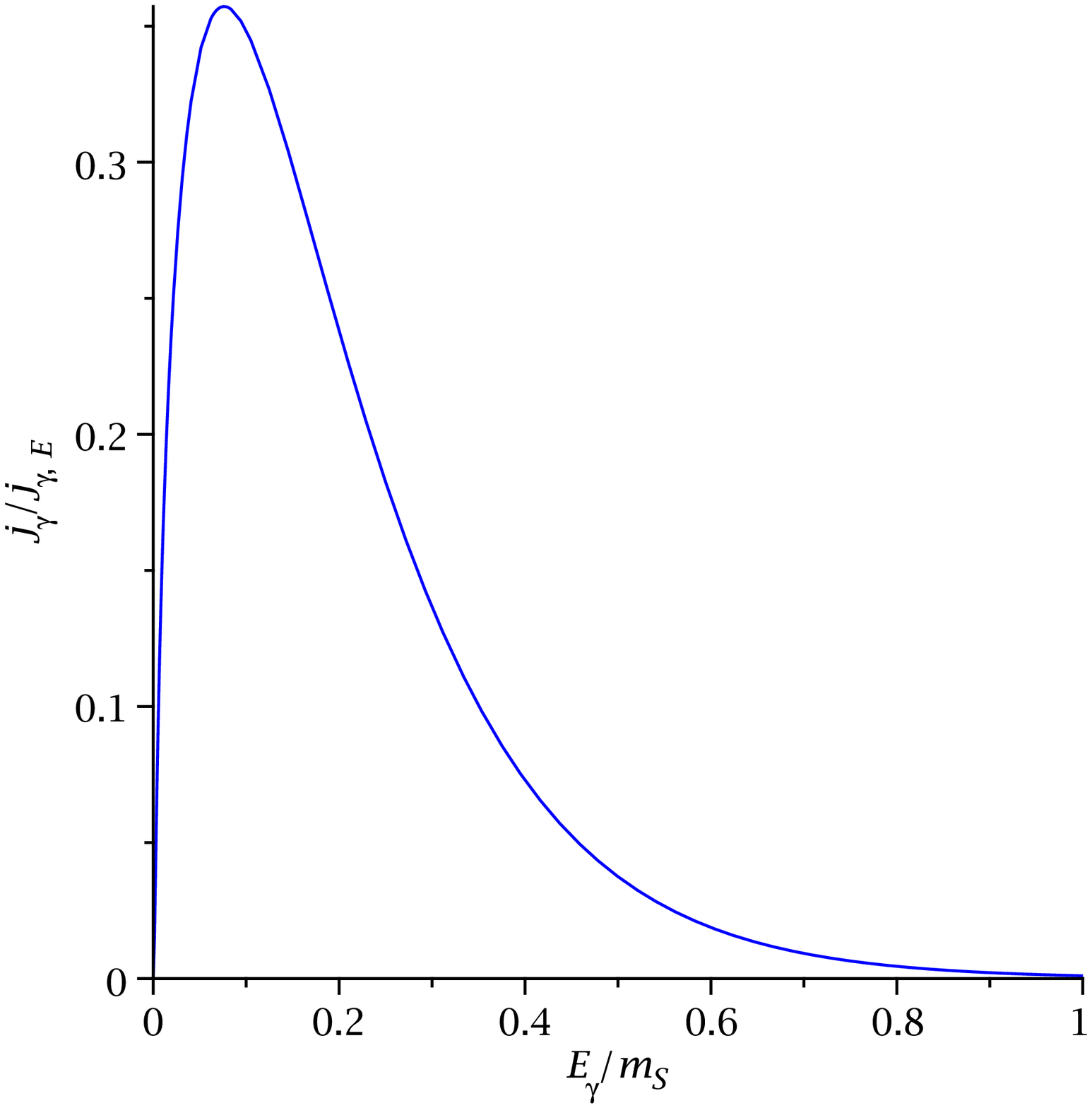}}
\end{center}
\caption{\it The photon flux $j_\gamma$ in the continuous spectrum for 
$m_h=160\,\mathrm{GeV}$, $m_S=120\,\mathrm{GeV}$, $\eta^2=0.1$ and 
$E_\gamma<120\,\mathrm{GeV}$ in units of the EGRET flux.}
\label{fig:phot160}     
\end{figure}

The corresponding flux ratio for $m_S=120$ GeV, $m_h=160$ GeV, $\eta^2=0.1$ 
and $v\sigma_{SS}=3.87\times 10^{-24}\,\mathrm{cm}^3/\mathrm{s}$
is diplayed in figure \ref{fig:phot160}.

The fluxes in figures \ref{fig:phot115} and \ref{fig:phot160} 
assume that all cold dark matter is
in a singlet state, but do not take into account correlations
between $m_S$ and $m_h$ from thermal creation of singlets.
This topic will be addressed in the next section.

\section{Thermal creation and abundance estimates}
\label{sec:LWabundance}

A thorough and beautiful application of Lee-Weinberg theory \cite{LW} for 
abundance estimates of electroweak singlets has been given in \cite{mcd}.
We will revisit the subject for the particular range of masses and cross 
sections of interest to us. The application of Lee-Weinberg theory in
reference \cite{mcd} used $\Omega_{dm}=1$ while the application for light 
electroweak singlets in \cite{cliff} used $\Omega_{dm}=0.6$. Here we use 
$\Omega_{dm}=0.2$ \cite{PDG}. 

In the absence of more detailed model assumptions about singlet
generating interactions behind the Higgs portal or co-annihilations, 
the effect of thermal creation of electroweak singlets below the
electroweak phase transition is taken into account through a thermal
production term in the rate equation
\begin{equation}\label{eq:nrateLW0}
\frac{d}{dt}(na^3)=\dot{N}_{thermal}-\langle\sigma v\rangle
n^2 a^3,
\end{equation}
where $a(t)$ is the scale factor in the Robertson-Walker metric.
The thermal production term is determined from the equilibrium
requirement $d(na^3)/dt=0$ \cite{LW,steigman},
\begin{equation}\label{eq:Nthermal}
\dot{N}_{thermal}=\langle\sigma v\rangle n_0^2 a^3,
\end{equation}
where $n_0$ is the thermal equilibrium density for temperature $T$.
The resulting rate equations during radiation domination is 
\begin{equation}\label{eq:nrateLW}
\frac{dn}{dt}+\frac{3n}{2t}=-\langle\sigma v\rangle 
\left(n^2-n_0^2\right).
\end{equation}

Radiation domination also yields
\begin{equation}\label{eq:tTrel1}
t=\frac{b}{T^2},
\end{equation}
and this is used to rewrite equation (\ref{eq:nrateLW}) in the form
\begin{equation}\label{eq:nrateLW2}
\frac{d}{dT}\frac{n}{T^3}=2b\langle\sigma v\rangle 
\frac{n^2-n_0^2}{T^6}.
\end{equation}
Equation (\ref{eq:tTrel1}) follows from the relations for the energy 
density in radiation for $t_{inflation}\ll t<t_{eq}$
($t_{eq}\simeq 2.4\times 10^{12}\,\mathrm{s}$ is the time of
 radiation-matter equality),
\begin{equation}\label{eq:rhogamma}
\varrho_\gamma=g_\ast(T)\frac{\pi^2(k_B T)^4}{30(\hbar c)^3}
=\frac{3m_{Planck}^2 c}{4\hbar t^2}.
\end{equation}
Here we use the reduced Planck mass 
$m_{Planck}=(\hbar c/8\pi G_N)^{1/2}$.
We have $g_\ast(T)=91.5$ for $m_b<T<m_W$. The parameter $b$ is
\begin{eqnarray*}
b&=&\frac{3\hbar m_{Planck}c^2}{\pi k_B^2}\sqrt{\frac{5}{2g_\ast(T)}}
\\ 
&=&2.53\times 10^{-7}k_B^{-2}\,\mathrm{s}\,\mathrm{GeV}^2
\\ 
&=&3.41\times 10^{19}\,\mathrm{s}\,\mathrm{K}^2.
\end{eqnarray*}
The analytic approximation proposed by Lee and Weinberg \cite{LW}
uses the equilibrium density
\begin{eqnarray*}
&&n_0(T)=\frac{1}{2\pi^2(\hbar c)^3}
\\
&&\times
\int_0^\infty dK\frac{(K+mc^2)\sqrt{K(K+2mc^2)}}{
\exp\left[(K+mc^2)/k_B T\right]-1}
\end{eqnarray*}
until a freeze-out temperature $T_f$ is reached with
\begin{equation}\label{eq:LWcond1}
\left.\frac{d}{dT}\frac{n_0(T)}{T^3}\right|_{T=T_f}
=2b\langle\sigma v\rangle \frac{n_0^2(T_f)}{T_f^6}.
\end{equation}
This temperature usually turns out to satisfy $T_f\lesssim 0.05m_S$,
such that the non-relativistic limit of $n_0(T)$,
\[
n_0(T)=\left(\frac{1}{\hbar}\sqrt{\frac{m_S k_B T}{2\pi}}\right)^3
\exp\left(-\frac{m_S c^2}{k_B T}\right),
\]
can be used in the evaluation of the Lee-Weinberg condition
(\ref{eq:LWcond1}). 
The solution is then extended for $T<T_f$ using
domination of the annihilation term in the rate equation
(\ref{eq:nrateLW0}),
\begin{equation}\label{eq:raterad}
\frac{dn}{dt}+\frac{3n}{2t}=-\langle\sigma v\rangle n^2,\quad
t_f<t<t_{eq}
\end{equation}
\begin{equation}\label{eq:ratelate}
\frac{dn}{dt}+\frac{3n}{a}\frac{da}{dt}
=-\langle\sigma v\rangle n^2,\quad
t>t_{eq},
\end{equation}
with the initial condition $n(t_f)=n_0(T_f)$.
Here $a=a(t)$ is the scale factor in the Robertson-Walker line element.
One might use the following set of equations for $t>t_{eq}$,
\[ 
\frac{dn}{dt}+\frac{2n}{t}=-\langle\sigma v\rangle n^2,\quad
t_{eq}<t<t_\Lambda,
\]
\[ 
\frac{dn}{dt}+\frac{2n}{\tau_\Lambda}\coth\left(\frac{t}{\tau_\Lambda}\right)
=-\langle\sigma v\rangle n^2,\quad t>t_\Lambda,
\]
with the time constant 
(for $\Lambda=0.76\varrho_c=4.27\,\mathrm{keV}/\mathrm{cm}^3$)
\[
\tau_\Lambda=2m_{Planck}\sqrt{\frac{c}{3\hbar\Lambda}}=3.23
\times 10^{17}\,\mathrm{s}, 
\]
and $t_\Lambda$
following from
\begin{eqnarray*}
1+z_\Lambda&\equiv&\frac{a(t_0)}{a(t_\Lambda)}
=\left(\frac{\sinh(t_0/\tau_\Lambda)}{\sinh(t_\Lambda/\tau_\Lambda)}\right)^{2/3}
\\
&=&\left(\frac{\Omega_\Lambda}{\Omega_m}\right)^{1/3}=1.47,
\end{eqnarray*}
see e.g. the appendix to reference \cite{rmk}.
However, we will see from the solution of equation (\ref{eq:raterad}) that
$n(t_{eq})\langle\sigma v\rangle t_{eq}<10^{-9}\ll 1$, see equation
(\ref{eq:latecold}) below. Therefore
both $n(t)\langle\sigma v\rangle t\ll 1$ and
$n(t)\langle\sigma v\rangle\tau_\Lambda\tanh(t/\tau_\Lambda)\ll 1$
for $t\ge t_{eq}$, i.e. the expansion term dominates strongly over the
annihilation term for $t>t_{eq}$. This yields standard cold dark matter
evolution for late times,
\[
n(t)=n(t_{eq})\left(\frac{a(t_{eq})}{a(t)}\right)^3,\quad
t>t_{eq}.
\]

We define $\xi=m_Sc^2/k_BT_f$. The Lee-Weinberg condition (\ref{eq:LWcond1}) 
takes the following form
\begin{eqnarray}\nonumber
\exp(\xi)&=&
\frac{2bk_B^2 m_S c^2}{(\sqrt{2\pi}\hbar c)^3}
\langle\sigma v\rangle\frac{\sqrt{\xi}}{\xi-1.5}
\\ \nonumber
&=&4.18\times 10^{11}
\frac{\langle\sigma v\rangle}{10^{-24}\,\mathrm{cm}^3/\mathrm{s}}
\\ \label{eq:LWcond2}
&&\times
\frac{m_S c^2}{100\,\mathrm{GeV}}\frac{\sqrt{\xi}}{\xi-1.5}.
\end{eqnarray}

We are interested in a perturbative Higgs portal $\eta^2\lesssim 0.1$
and weak scale singlet masses. Our previous results on cross sections
then imply that the factor
$(\langle\sigma v\rangle/10^{-24}\,\mathrm{cm}^3\,\mathrm{s}^{-1})
\times (m_S/100\,\mathrm{GeV})$ should be in the range between 0.1
and 10. This will yield values for $\xi$ between 20 and 30. Thermal 
theories of particle creation using equation (\ref{eq:Nthermal}) 
generically predict that particles will remain thermal until the 
temperature has dropped to a value well below their mass threshold.

Integration of equation (\ref{eq:raterad}) yields
\begin{eqnarray}\nonumber
n(t_{eq})&=&\left[
\frac{1}{n(t_f)}\left(\frac{t_{eq}}{t_f}\right)^{3/2}
+2\langle\sigma v\rangle t_f\right.
\\ \nonumber
&&\times\left.
\left(\left(\frac{t_{eq}}{t_f}\right)^{3/2}-\frac{t_{eq}}{t_f}\right)
\right]^{-1}
\\ \label{eq:nteq}
&\simeq&\frac{n(t_f)}{1+2n(t_f)\langle\sigma v\rangle t_f}
\left(\frac{t_f}{t_{eq}}\right)^{3/2},
\end{eqnarray}
where we used that freeze-out temperatures following from 
(\ref{eq:LWcond2}) will at least be a few GeV for the parameter range
of interest here, and therefore $(t_{eq}/t_f)^{1/2}>10^9$.


The relation between temperature and time and the definition of $\xi$
imply
\begin{eqnarray}\nonumber
t_f&=&\frac{b}{T_f{}^2}=\frac{bk_B{}^2}{m_S{}^2 c^4}\xi^2
\\ \label{eq:ntf}
&=&2.53\times 10^{-11}\xi^2\,\mathrm{s}
\times\left(\frac{100\,\mathrm{GeV}}{m_Sc^2}\right)^2.
\end{eqnarray}
This yields
\begin{eqnarray*}
\langle\sigma v\rangle t_f&=&2.53\times 10^{-35}\xi^2\,\mathrm{cm}^{3}
\\
&&\times\left(\frac{100\,\mathrm{GeV}}{m_Sc^2}\right)^2
\times\frac{\langle\sigma v\rangle}{10^{-24}\,\mathrm{cm}^{3}/\mathrm{s}}.
\end{eqnarray*}
On the other hand, the density is with (\ref{eq:LWcond2})
\begin{eqnarray*}
n(t_f)&=&\left(\frac{m_Sc}{\sqrt{2\pi\xi}\hbar}\right)^3\exp(-\xi)
\\
&=&\frac{m_S{}^2 c^4}{2bk_B{}^2\langle\sigma v\rangle}\frac{\xi-1.5}{\xi^2}
\\
&=&1.98\times 10^{34}\frac{\xi-1.5}{\xi^2}\,\mathrm{cm}^{-3}
\\
&&\times\left(\frac{m_Sc^2}{100\,\mathrm{GeV}}\right)^2
\times\frac{10^{-24}\,\mathrm{cm}^{3}/\mathrm{s}}{\langle\sigma v\rangle},
\end{eqnarray*}
and therefore 
\begin{equation}\label{eq:prodxi}
2n(t_f)\langle\sigma v\rangle t_f=\xi-1.5.
\end{equation}
This equation also implies
\begin{eqnarray}\nonumber
n(t_{eq})\langle\sigma v\rangle t_{eq}
&=&\frac{n(t_f)\langle\sigma v\rangle t_f}{1+
2n(t_f)\langle\sigma v\rangle t_f}\sqrt{\frac{t_f}{t_{eq}}}
\\ \label{eq:latecold}
&=&\frac{\xi-1.5}{2\xi-1}\sqrt{\frac{t_f}{t_{eq}}}<10^{-9},
\end{eqnarray}
and therefore the annihilation term is much smaller than the
expansion term for $t\ge t_{eq}$, $d(na^3)/dt=0$,
\begin{equation}\label{eq:nt0}
n(t_0)=n(t_{eq})z_{eq}^{-3},
\end{equation}
where
\[
z_{eq}\equiv\frac{a(t_0)}{a(t_{eq})}-1\simeq\frac{a(t_0)}{a(t_{eq})}.
\]
We use $z_{eq}=3000$ for numerical work.

Equations (\ref{eq:nteq}),  (\ref{eq:ntf}),  (\ref{eq:prodxi}), and 
(\ref{eq:nt0}) yield the current energy density in one singlet species,
\begin{eqnarray}\nonumber
\varrho_S^{(1)}&=&n(t_{eq})m_Sc^2
\\ \nonumber
&=&\frac{2\xi-3}{2\xi-1}\xi
\frac{k_B\sqrt{b}}{2\langle\sigma v\rangle 
t_{eq}{}^{3/2} z_{eq}{}^{3}}
\\ \label{eq:rhoS1}
&\simeq&\frac{2\xi-3}{2\xi-1}\xi\times
\frac{2.51\,\mathrm{eV}/\mathrm{cm}^3}{
\langle\sigma v\rangle/10^{-24}\,\mathrm{cm}^3/\mathrm{s}}.
\end{eqnarray}
However, we can have an $O(N)$ symmetric $N$-plet of electroweak singlets 
of mass $m_S$, such that the current energy density in singlets is
\begin{equation}\label{eq:rhoSN}
\varrho_S\simeq\frac{2\xi-3}{2\xi-1}\xi\times
\frac{N\times 2.51\,\mathrm{eV}/\mathrm{cm}^3}{
\langle\sigma v\rangle/10^{-24}\,\mathrm{cm}^3/\mathrm{s}}.
\end{equation}

\setcounter{table}{1}

The condition $\varrho_S=\varrho_{dm}=1.106\,\mathrm{keV}/\mathrm{cm}^3$
then determines $\xi$ in terms of $\langle\sigma v\rangle/N$, and 
substitution in equation (\ref{eq:LWcond2}) then relates $m_S$, $m_h$, 
and $N$. We report the resulting singlet masses in the mass range 
$m_W<m_S<1\,\mathrm{TeV}$ for $\eta^2=0.1$ and for $N=1$ or $N=10$ 
in table 1. We also report the corresponding freeze out temperatures.
The annihilation cross sections are given in the form
\[
x_\sigma\equiv\frac{v\sigma_{SS}}{10^{-24}\,\mathrm{cm}^3/
\mathrm{s}}.
\]

\begin{table}

\begin{tabular}{|l|l|l|}
       & $m_h=115$ GeV & $m_h=160$ GeV \\ \hline 
$N=1$  & $m_S=879$ GeV & $m_S=883$ GeV \\
       & $T_f=35.9$ GeV & $T_f=36.1$ GeV\\ 
       & $x_\sigma=0.0531$ & $x_\sigma=0.0531$\\ \hline 
$N=10$ & $m_S=273$ GeV & $m_S=285$ GeV \\
       & $T_f=10.7$ GeV & $T_f=11.1$ GeV\\ 
       & $x_\sigma=0.557$ & $x_\sigma=0.558$\\ \hline 
\end{tabular}

\vspace*{2mm}
Table 1: {\it Singlet masses in the mass range
$m_W<m_S<1\,\mathrm{TeV}$ which satisfy
the Lee-Weinberg condition (\ref{eq:LWcond2}) for $\eta^2=0.1$,
$m_h=115\,\mathrm{GeV}$ or $m_h=160\,\mathrm{GeV}$, 
and $N=1$ or $N=10$.}

\end{table}

Table 2 shows solutions in the mass range $m_W<m_S<1\,\mathrm{TeV}$ for 
very weak coupling $\eta^2=0.01$.

\begin{table}

\begin{tabular}{|l|l|l|}
       & $m_h=115$ GeV & $m_h=160$ GeV \\ \hline 
$N=1$  & $m_S=293$ GeV & $m_S=304$ GeV \\
       & $T_f=12.6$ GeV & $T_f=13.0$ GeV\\ 
       & $x_\sigma=0.0505$ & $x_\sigma=0.0506$\\ \hline 
$N=10$ & n/a & $m_S=111$ GeV \\
       &  & $T_f=4.47$ GeV\\ 
       &  & $x_\sigma=0.540$\\ \hline 
\end{tabular}

\vspace*{2mm}
Table 2: {\it Singlet masses in the mass range
$m_W<m_S<1\,\mathrm{TeV}$ which satisfy
the Lee-Weinberg condition (\ref{eq:LWcond2}) for $\eta^2=0.01$,
$m_h=115\,\mathrm{GeV}$ or $m_h=160\,\mathrm{GeV}$, 
and $N=1$ or $N=10$.}

\end{table}

The cross sections in table 2 for given $m_h$ and $N$ are similar to
the cross sections in table 1, in spite of the weaker coupling. This
is due to the fact that the corresponding singlet masses in table 2 are 
smaller and much closer to the peaks in the cross section.

 For the flux calculations for $N>1$, we have to rescale the flux
(\ref{eq:flux1c}) by a factor $1/N$, because the density factor $n^2$
for each annihilating species is now suppressed $\propto N^{-2}$, but 
there are $N$ annihilating species of singlets of mass $m_S$.
Therefore we find with equations (\ref{eq:flux1c}) and (\ref{eq:lars1}) 
\begin{eqnarray}\nonumber
j_\gamma&=&\frac{1.95\times 10^{20}}{N}\times
\left(\frac{\mathrm{GeV}}{m_S}\right)^2
\\ \nonumber
&&\times\frac{d\mathcal{N}_\gamma(E_\gamma)}{dE_\gamma}
\frac{v\sigma_{SS}}{\mathrm{cm}^5\,\mathrm{sr}}
\\ \nonumber
&=&\frac{8.19\times 10^{19}}{N}\times
\left(\frac{\mathrm{GeV}}{m_S}\right)^2
\\ \label{eq:fluxgamma}
&&\times\frac{\sqrt{m_S}\exp(-8E_\gamma/m_S)}{E_\gamma{}^{1.5}+0.00014m_S{}^{1.5}}
\frac{v\sigma_{SS}}{\mathrm{cm}^5\,\mathrm{sr}}.
\end{eqnarray}
  
The parameter 
$(v\sigma_{SS}/N)/(10^{-24}\,\mathrm{cm}^3/\mathrm{s})
=x_\sigma/N$ varies only in the range 
$5.05\times 10^{-2}\le x_\sigma/N\le 5.58\times 10^{-2}$ for the 
solutions in tables 1 and 2. However, $j_\gamma$ scales with 
$m_S{}^{-1.5}$, and therefore the low-mass solutions from table 2 
yield higher flux than the solutions from table 1. We will give 
results for $j_\gamma$ in the case $\eta^2=0.1$, $N=10$, $m_h=160$ 
GeV and $m_S=285$ GeV, and also in the case $\eta^2=0.01$, $N=10$, 
$m_h=160$ GeV and $m_S=111$ GeV.

\begin{figure}[htb]
\begin{center}
\scalebox{0.38}{\includegraphics{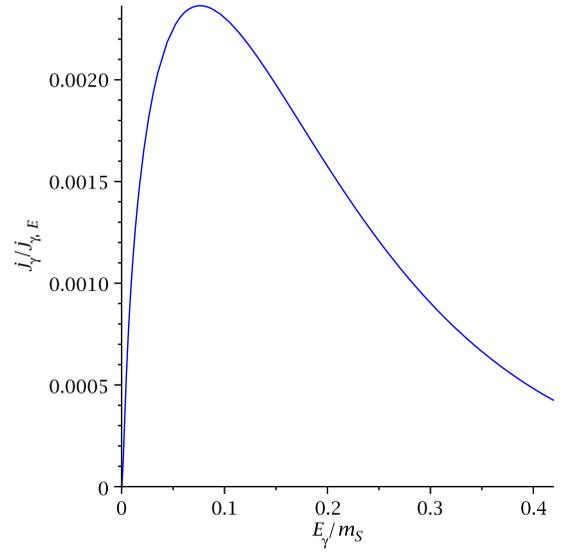}}
\end{center}
\caption{\it The photon flux $j_\gamma$ in the continuous spectrum for 
$\eta^2=0.1$, $N=10$, $m_h=160\,\mathrm{GeV}$, $m_S=285\,\mathrm{GeV}$ and 
$E_\gamma<120\,\mathrm{GeV}$ in units of the EGRET flux.}
\label{fig:flux160285}     
\end{figure}

\begin{figure}[htb]
\begin{center}
\scalebox{0.38}{\includegraphics{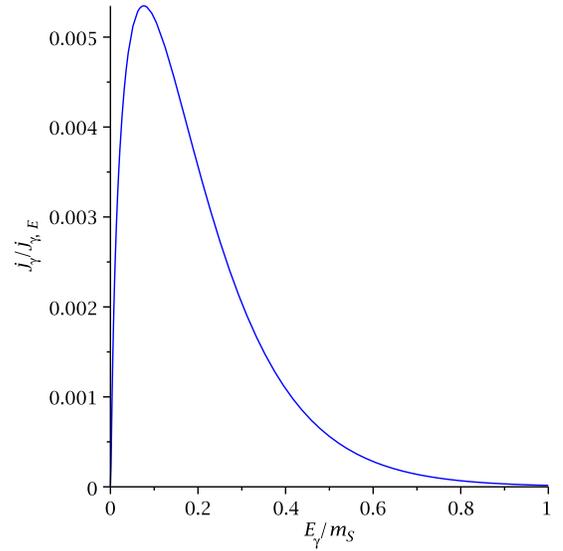}}
\end{center}
\caption{\it The photon flux $j_\gamma$ in the continuous spectrum for 
$\eta^2=0.01$, $N=10$, $m_h=160\,\mathrm{GeV}$, $m_S=111\,\mathrm{GeV}$ and 
$E_\gamma<111\,\mathrm{GeV}$ in units of the EGRET flux.}
\label{fig:flux160111} 
\end{figure}

\begin{figure}[htb]
\begin{center}
\scalebox{0.38}{\includegraphics{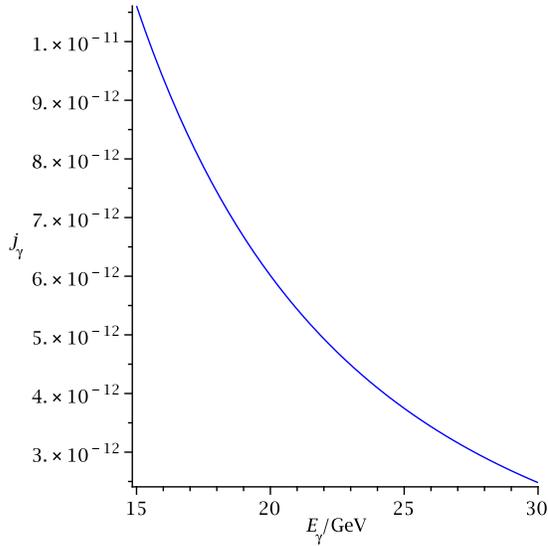}}
\end{center}
\caption{\it The photon flux $j_\gamma$ in units of 
$(\mathrm{GeV}\,\mathrm{cm}^2\,\mathrm{s}\,\mathrm{sr})^{-1}$
for $\eta^2=0.1$, $N=10$, $m_h=160\,\mathrm{GeV}$, $m_S=285\,\mathrm{GeV}$ 
and $E_\gamma$ between $15\,\mathrm{GeV}$ and $30\,\mathrm{GeV}$. 
The maximum in figure \ref{fig:flux160285} corresponds to 
$E_\gamma\simeq 22\,\mathrm{GeV}$.}
\label{fig:j160285}     
\end{figure}

\begin{figure}[htb]
\begin{center}
\scalebox{0.38}{\includegraphics{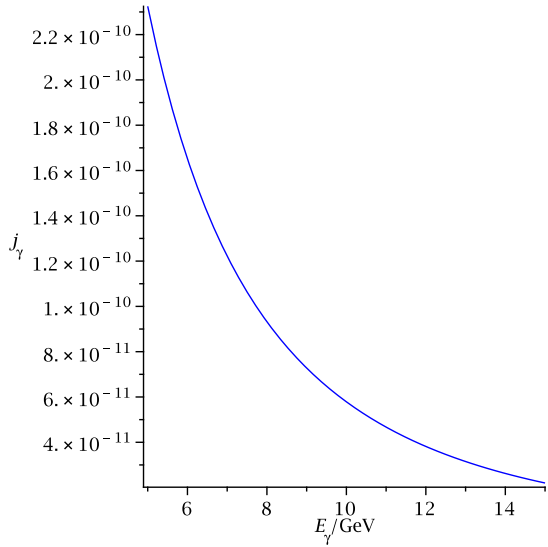}}
\end{center}
\caption{\it The photon flux $j_\gamma$ in units of 
$(\mathrm{GeV}\,\mathrm{cm}^2\,\mathrm{s}\,\mathrm{sr})^{-1}$
for $\eta^2=0.01$, $N=10$, $m_h=160\,\mathrm{GeV}$, $m_S=111\,\mathrm{GeV}$ 
and $E_\gamma$ between $5\,\mathrm{GeV}$ and $15\,\mathrm{GeV}$. 
The maximum in figure \ref{fig:flux160111} corresponds to 
$E_\gamma\simeq 9\,\mathrm{GeV}$.}
\label{fig:j160111}     
\end{figure}

The expected contribution to the photon flux below 120 GeV in units of the
EGRET flux is shown in figures \ref{fig:flux160285} and \ref{fig:flux160111}
for the two cases. The corresponding number of photons per 
$\mathrm{GeV}\cdot\mathrm{cm}^2\cdot\mathrm{s}\cdot\mathrm{sr}$ 
is given in figures \ref{fig:j160285} and \ref{fig:j160111}, respectivley.

\begin{figure}[htb]
\begin{center}
\scalebox{0.38}{\includegraphics{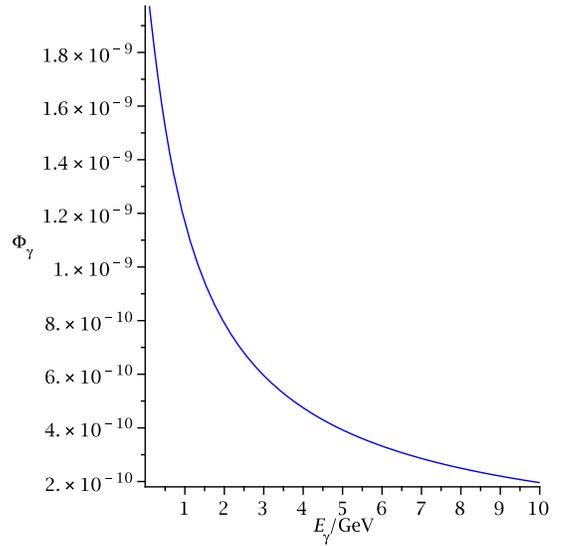}}
\end{center}
\caption{\it The integrated photon flux $\Phi_\gamma$ above energy 
$E_\gamma$ in units of $(\mathrm{cm}^2\,\mathrm{s}\,\mathrm{sr})^{-1}$
for $\eta^2=0.1$, $N=10$, $m_h=160\,\mathrm{GeV}$, $m_S=285\,\mathrm{GeV}$ 
and $E_\gamma$ between $100\,\mathrm{MeV}$ and $10\,\mathrm{GeV}$.}
\label{fig:Phi160285}     
\end{figure}

\begin{figure}[htb]
\begin{center}
\scalebox{0.38}{\includegraphics{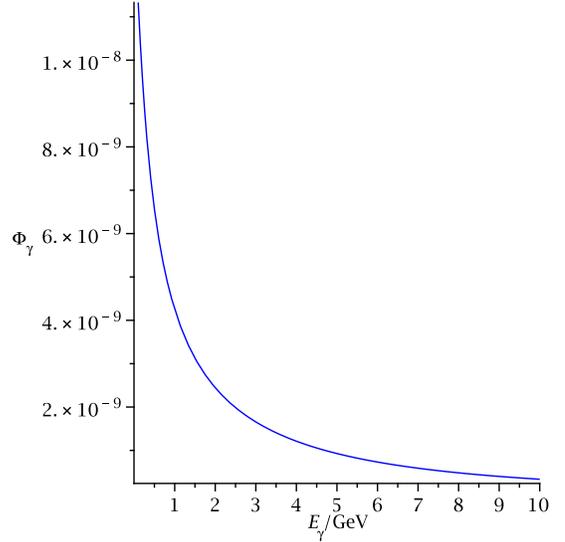}}
\end{center}
\caption{\it The integrated photon flux $\Phi_\gamma$ above energy 
$E_\gamma$ in units of $(\mathrm{cm}^2\,\mathrm{s}\,\mathrm{sr})^{-1}$
for $\eta^2=0.01$, $N=10$, $m_h=160\,\mathrm{GeV}$, $m_S=111\,\mathrm{GeV}$ 
and $E_\gamma$ between $100\,\mathrm{MeV}$ and $10\,\mathrm{GeV}$.}
\label{fig:Phi160111} 
\end{figure}

The integrated photon flux 
\[
\Phi_\gamma(E_\gamma)=\int_{E_\gamma}^\infty dE j_\gamma(E)
\]
in the two cases is shown in figures \ref{fig:Phi160285}
and \ref{fig:Phi160111}.

 For an instrument with an effective area of $8000\,\mathrm{cm}^2$ and a
field of view of $2.4\,\mathrm{sr}$, comparable to the Large Area 
Telescope on GLAST, the photon flux $j_\gamma$ in figure \ref{fig:j160111} 
corresponds to an annual rate of 470 photons with energies between 
$5\,\mathrm{GeV}<E_\gamma<15\,\mathrm{GeV}$ from dark matter annihilation 
on a diffuse astrophysical background of 90,000 photons per year. The flux
in figure \ref{fig:j160285} corresponds to an annual rate of 48 photons 
with energies between $15\,\mathrm{GeV}<E_\gamma<30\,\mathrm{GeV}$ from 
dark matter annihilation on a diffuse astrophysical background of 22500 
photons per year. There are two characteristic features which would help 
to identify $j_\gamma$ as an excess due to dark matter annihilation. The 
excess would extend over an energy range of order 10 GeV, and it would be 
correlated with the galactic halo.

\section{Conclusions}\label{sec:conc}

We have considered perturbatively coupled electroweak singlet dark 
matter in the intermediate mass range $m_W<m_S<1\,\mathrm{TeV}$. The 
leading order annihilation cross section of the singlets is enhanced and 
varies strongly due to proximity to the $W$, $Z$ and Higgs peaks. The 
product $v\sigma_{SS}$ is of order 
$\eta^2\times 10^{-23}\,\mathrm{cm}^3/\mathrm{s}$ for $m_S$ close to 
$m_h$, i.e. it can substantially exceed standard estimates of dark matter 
annihilation cross sections even for perturbative singlet-Higgs coupling. 

 For singlet masses above the $SS\to WW$ threshold and electroweak 
strength coupling $\eta^2\simeq 0.1$, the Lee-Weinberg condition and the 
requirement $\Omega_S=\Omega_{dm}$ push $m_S$ to high mass values around 
900 GeV if there is only one singlet state. However, if there is an 
$N$-plet of electroweak singlets or if the coupling is weaker, 
$\eta^2\lesssim 0.01$, lower singlet mass values can be achieved, and the 
annihilation signal from the continuous $\gamma$ ray spectrum can reach a 
level of several per mil of the EGRET diffuse $\gamma$ ray signal for 
photon energies $E_\gamma\sim 0.08m_S$. This excess contribution over the
expected cosmological background would appear typically over an energy 
range of order 10 GeV, and its dark matter signature would be its 
correlation with the galactic halo. 

The flux reported by EGRET was the diffuse $\gamma$ ray flux after 
subtraction of then known or expected galactic components. Reduction of 
the diffuse ``excess'' $\gamma$ ray flux due to subtraction of a larger 
component from interstellar gas and standard extragalactic sources 
increases the relative importance of an annihilation signal for any 
possible excess signal and improves detectability.
The diffuse $\gamma$ ray flux will be measured with higher sensitivity
and precision in the near future by the Large Area Telescope aboard the 
GLAST satellite. This will also cover a larger energy range up to 300 GeV.

The minimal dark matter models considered here include four basic parameters,
the singlet mass $m_S$, the number of singlet states $N$, the singlet-Higgs
coupling $\eta$, and the Higgs mass $m_h$. The assumption that electroweak
singlets of mass $m_S$ provide the dark matter in the universe relates these
parameters through Lee-Weinberg theory. Measuring the Higgs mass at the Tevatron
or the LHC will reduce the number of free parameters in this class of minimal
dark matter models to two, or maybe even to only one free parameter
if a missing energy signal can be used to constrain a combination of $\eta$, 
$N$ and $m_S$. The anticipated smallness of singlet annihilation signals in 
cosmic $\gamma$ rays indicates that observation of the Higgs particle at 
the Tevatron or the LHC may be needed for a successful search for a singlet
annihilation signal in cosmic rays.

\section*{Acknowledgement}

This work was supported by NSERC Canada. RD thanks Freddy Cachazo,
Rob Myers and Tom Waterhouse for interesting discussions, and gratefully 
acknowledges the hospitality of the Perimeter Institute for Theoretical 
Physics. We also thank the referee for constructive criticism of a first 
draft of this manuscript.


\begin{thebibliography}{99}

\bibitem{axion} 
J. Preskill, M.B. Wise, F. Wilczek,
Phys. Lett. B 120 (1983) 127;
L.F. Abbott, P. Sikivie,
Phys. Lett. B 120 (1983) 133;
M. Dine, W. Fischler,
Phys. Lett. B 120 (1983) 137.

\bibitem{neutralino} 
S. Weinberg, Phys. Rev. Lett. 50 (1983) 387;
H. Goldberg, Phys. Rev. Lett. 50 (1983) 1419;
J.R. Ellis, J.S. Hagelin, D.V. Nanopoulos, K.A. Olive,
M. Srednicki, Nucl. Phys. B 238 (1984) 453.

\bibitem{rocky1} 
E.W. Kolb, R. Slansky, Phys. Lett. B 135 (1984) 378.

\bibitem{dilaton} 
Y.M. Cho, Phys. Rev. D 41 (1990) 2462;
R. Dick, Mod. Phys. Lett. A 12 (1997) 47;
Fortschr. Phys. 45 (1997) 537;
Y.M. Cho, Y.Y. Keum, Mod. Phys. Lett. A 13 (1998) 109;
Class. Quantum Grav. 15 (1998) 907;
Y.M. Cho, J.H. Kim, {\it Dilaton as a dark matter 
candidate and its detection}, arXiv:0711.2858 [gr-qc].

\bibitem{rocky2} 
D.J. Chung, E.W. Kolb, A. Riotto, 
Phys. Rev. Lett. 81 (1998) 4048; Phys. Rev. D 59 (1999) 023501.

\bibitem{KT} 
V.A. Kuzmin, I.I. Tkachev, Phys. Rev. D 59 (1999) 123006;
Phys. Rep. 320 (1999) 199.

\bibitem{lev} 
L. Kofman, A.D. Linde, A.A. Starobinsky, 
Phys. Rev. Lett. 73 (1994) 3195; Phys. Rev. D 56 (1997)
3258.

\bibitem{dan} 
G. Bertone, D. Hooper, J. Silk,
Phys. Rep. 405 (2005) 279.

\bibitem{dan2} 
D. Hooper, T. Plehn, Phys. Lett. B 562 (2003) 18.

\bibitem{bottino} 
A. Bottino, F. Donato, N. Fornengo, S. Scopel,
Phys. Rev. D 68 (2003) 043506.

\bibitem{zee} 
V. Silveira, A. Zee, Phys. Lett. B 161 (1985) 136.

\bibitem{mcd}
J. McDonald, Phys. Rev. D 50 (1994) 3637.

\bibitem{cliff} 
C.P. Burgess, M. Pospelov, T. ter Veldhuis,
Nucl. Phys. B 619 (2001) 709.

\bibitem{mura}
H. Davoudiasl, R. Kitano, T. Li, H. Murayama,
Phys. Lett. B 609 (2005) 117.

\bibitem{frank}
B. Patt, F. Wilczek, {\it Higgs-field portal into
hidden sectors}, hep-ph/0605188.

\bibitem{zee2} 
D.E. Holz, A. Zee, Phys. Lett. B 517 (2001) 239.

\bibitem{maxim}
C. Bird, R. Kowalewski, M. Pospelov, Mod. Phys. Lett.
A 21 (2006) 457.

\bibitem{hitoshi}
H. \nolinebreak Murayama, \nolinebreak 
{\it Physics \nolinebreak beyond \nolinebreak the Standard Model and
dark matter}, arXiv:0704.2276 [hep-ph].

\bibitem{cirelliA}
M. Cirelli, N. Fornengo, A. Strumia, 
Nucl. Phys. B 753 (2006) 178;
M. Cirelli, A. Strumia, M. Tamburini,
Nucl. Phys. B 787 (2007) 152.

\bibitem{cirelli}
M. Cirelli, R. Franceschini, A. Strumia,
Nucl. Phys. B 800 (2008) 204.

\bibitem{mcd2}
J. McDonald, N. Sahu, JCAP 0806 (2008) 026.

\bibitem{BR} 
O. Bertolami, R. Rosenfeld, 
{\it The Higgs portal and a unified model for dark energy 
and dark matter}, arXiv:0708.1784 [hep-ph].

\bibitem{CNW} 
W.F. Chang, J.N. Ng, J.M.S. Wu,
Phys. Rev. D 74 (2006) 095005; Phys. Rev. D 75 (2007)
115016; {\it Phenomenology from a U(1) gauged hidden
sector}, arXiv:0706.2345 [hep-ph].

\bibitem{hermann}
K.A. Meissner, H. Nicolai, Phys. Lett. B 648 (2007) 312.

\bibitem{EQ}
J.R. Espinosa, M. Quir\'os, Phys. Rev. D 76 (2007) 076004.

\bibitem{PRS}
S. Profumo, M.J. Ramsey-Musolf, G. Shaughnessy,
JHEP 08 (2007) 010.

\bibitem{DR}
A. Datta, A. Raychaudhuri, Phys. Rev. D 57 (1998) 2940.

\bibitem{dav1}
H. Davoudiasl, T. Han, H.E. Logan,
Phys. Rev. D 71 (2005) 115007

\bibitem{SW}
R. Schabinger, J.D. Wells,
Phys. Rev. D 72 (2005) 093007.

\bibitem{john}
John March-Russell, Stephen M. West, Daniel Cumberbatch, Dan Hooper,
{\it Heavy dark matter through the Higgs portal},
arXiv:0801.3440 [hep-ph].

\bibitem{BDK}
P. Blasi, R. Dick, E.W. Kolb,
Astropart. Phys. 18 (2002) 57.

\bibitem{lars2}
P. Ullio, L. Bergstr\"om, J. Edsj\"o, C. Lacey,
Phys. Rev. D 66 (2002) 123502.


\bibitem{bergstrom}
L. Bergstr\"om, P. Ullio, J.H. Buckley,
Astropart. Phys. 9 (1998) 137.

\bibitem{sarkar}
N.W. Evans, F. Ferrer, S. Sarkar,
Phys. Rev. D 69 (2004) 123501.

\bibitem{nfw}
J.F. Navarro, C.S. Frenk, S.D.M. White,
Mon. Not. R. Astron. Soc. 275 (1995) 720;
Astrophys. J. 462 (1996) 563.

\bibitem{klypin}
A. Klypin, H. Zhao, R.S. Somerville,
Astrophys. J. 573 (2002) 597.

\bibitem{boyarski}
A. Boyarski, A. Neronov, O. Ruchayskiy, M. Shaposhnikov,
I. Tkachev, Phys. Rev. Lett. 97 (2006) 261302.
 
\bibitem{bmp}
{\it Higher Transcendental Functions Vol. 1}, 
ed. A. Erd\'{e}lyi (McGraw-Hill, New York 1953) pp. 31-32.

\bibitem{AS} 
{\it Hand\-book of Mathematical Functions}, 
eds. M. Abramowitz and I.A. Stegun
(United States National Bureau of Standards, 
10th printing, Washington 1972) pp. 1004-1005.

\bibitem{biebel}
O. Biebel, D. Milstead, P. Nason, B.R. Webber,
{\it Fragmentation functions in $e^+ e^-$
annihilations and lepton-nucleon DIS},
in \cite{PDG}.

\bibitem{hrs}
D. Bender et al. (HRS Collaboration),
Phys. Rev. D 31 (1985) 1.

\bibitem{lars1}
L. Bergstr\"om, J. Edsj\"o, P. Ullio,
Phys. Rev. Lett. 87 (2001) 251301.

\bibitem{KKT}
M. Kaplinghat, L. Knox, M.S. Turner,
Phys. Rev. Lett. 85 (2000) 3335.

\bibitem{BBM}
J.F. Beacom, N.F. Bell, G.D. Mack,
Phys. Rev. Lett. 99 (2007) 231301.

\bibitem{yuksel}
H. Y\"uksel, S. Horiuchi, J.F. Beacom, S. Ando,
Phys. Rev. D 76 (2007) 123506.

\bibitem{steve2}
S. Weinberg, {\it The Quantum Theory of Fields Vol. 1} 
(Cambridge University Press, Cambridge 1995) Sec. 3.7.

\bibitem{unit}
K. Griest, M. Kamionkowski, Phys. Rev. Lett. 64 (1990) 615;
L. Hui, Phys. Rev. Lett. 86 (2001) 3467.

\bibitem{hegra}
F.A. Aharonian, W. Wittek et al. (HEGRA Collaboration),
Astropart. Phys. 17 (2002) 459.

\bibitem{egret} 
P. Sreekumar et al., Astrophys. J. 494 (1998) 523.

\bibitem{disc1}
A.W. Strong, I.V. Moskalenko, O. Reimer,
Astrophys. J. 613 (2004) 962.

\bibitem{disc2}
F.W. Stecker, S.D. Hunter, D.A. Kniffen,
Astropart. Phys. 29 (2008) 25.

\bibitem{DGH}
J.F. Donoghue, E. Golowich, B.R. Holstein,
{\it Dynamics of the Standard Model}
(Cambridge University Press, Cambridge 1992) Sec. XV.2.

\bibitem{PDG}
W.-M. Yao et al. (Particle Data Group),
J. Phys. G 33 (2006) 1
and 2007 partial update for the 2008 edition.

\bibitem{teva}
M.P. Sanders, {\it Search for Low Mass SM Higgs at the Tevatron},
arXiv:0805.1248 [hep-ex];
A. Duperrin, {\it Review of Searches for Higgs Bosons and Beyond the
Standard Model Physics at the Tevatron}, arXiv:0805.3624 [hep-ex].

\bibitem{LW}
B.W. Lee, S. Weinberg, Phys. Rev. Lett. 39 (1977) 165.

\bibitem{steigman}
G. Steigman, Ann. Rev. Nucl. Part. Sci. 29 (1979) 313.

\bibitem{rmk}
M.G. Barnett, R. Dick, K.E. Wunderle, Mon. Not. R. Astron. Soc.
349 (2004) 1500.


\end{thebibliography}
\end{document}